\documentclass[onefignum,onetabnum]{siamart190516}

\pdfoutput=1

\input{./anc/shared}

\ifpdf
\hypersetup{
  pdftitle={Pressio},
  pdfauthor={F.~Rizzi, P.~J.~Blonigan, E.J.Parish, and K.~T.~Carlberg}
}
\fi

\begin{document}

\maketitle

\begin{abstract}
  This work introduces \pressio, an open-source project aimed at enabling
  leading-edge projection-based reduced order models (ROMs)
  for large-scale nonlinear dynamical systems in science and engineering.
  \pressio~provides model-reduction methods that can reduce both the number of
  spatial and temporal degrees of freedom for any dynamical system expressible
  as a system of parameterized ordinary differential equations (ODEs).
  We leverage this simple, expressive mathematical framework as a pivotal
  design choice to enable a minimal application programming interface (API)
  that is natural to dynamical systems.
  The core component of \pressio~is a C++11 header-only library that leverages
  generic programming to support applications with arbitrary data types
  and arbitrarily complex programming models. This is complemented with Python bindings
  to expose these C++ functionalities to Python users with negligible
  overhead and no user-required binding code.
  We discuss the distinguishing characteristics of \pressio~relative to
  existing model-reduction libraries,
  outline its key design features, describe how the user interacts with it,
  and present two test cases---including one with over 20 million degrees of
  freedom---that highlight the performance results of
  \pressio~and illustrate the breath of problems that can be addressed with
  it.
\end{abstract}

\begin{keywords}
	projection-based model reduction, Galerkin, LSPG, POD, SVD, sample mesh,
	hyper-reduction, scientific computing, object-oriented programming, generic
	programming, policy-based design, design by introspection, template
	metaprogramming, HPC, Kokkos, Trilinos, Python.
\end{keywords}

\begin{AMS}
65L05, 65L06, 65M12, 68U20, 68N15, 68W10, 68W15, 68N01, 76K05
\end{AMS}

\section{Introduction}
\label{sec:intro}

Computational science and engineering has become a fundamental driver of the
technological advancement of modern society, underpinning applications such as
scientific discovery, engineering design, vehicle control, infrastructure
monitoring, and risk assessment of engineered systems.  Computational science
and engineering and high-performance computing have a symbiotic relationship:
the constant improvement and availability of computing power drives the
ambition to model increasingly complex systems with increasing levels of
fidelity, which, in turn, motivates the pursuit of more powerful computing
platforms.

The present ``extreme-scale'' computing era---often referred to
as the dawn of exascale computing---is one of unprecedented access to
computational resources. In principle, this enables scientists and engineers
to tackle very complex problems that are
\textit{time critical} or \textit{many query} in nature.
Time-critical problems are those constrained by a limited wall-clock time, and
include fast-turnaround design, rapid path planning, and model predictive
control, which require simulations to execute on the order of hours, minutes,
or milliseconds, respectively.  In contrast, many-query problems are
characterized by a limited budget of computational core-hours. Many-query
problems pervade uncertainty quantification (UQ), as such applications
typically require thousands of model evaluations to characterize
adequately the effect of uncertainties in parameters and operating conditions
on the system response.  In the case of both time-critical and many-query
problems, if the system of interest is
computationally expensive to query---as in the case of high-fidelity models
for which a single run can consume days or weeks on a supercomputer---the
constraint on wall-clock time or core-hours can be impossible to satisfy.
In such scenarios, analysts often turn to surrogate models, which replace the
high-fidelity model with a lower-cost, lower-fidelity counterpart that can be
employed within the application to make satisfaction of the wall-clock-time or
core-hour constraint feasible. In this work, we consider the high-fidelity
model of interest to be a nonlinear dynamical system expressible a system of
parameterized ordinary differential equations (ODEs).


Broadly speaking, surrogate models can be classified under three categories,
namely (a) \textit{data fits}, which
construct an explicit mapping (e.g., using polynomials, Gaussian processes)
from the system's parameters (i.e., inputs) to
the system response of interest (i.e., outputs), (b) \textit{lower-fidelity
models}, which simplify the high-fidelity model (e.g., by coarsening the mesh,
employing a lower finite-element order, or neglecting physics), and (c)
\textit{projection-based reduced-order models (ROMs)}, which reduce the number
of degrees of freedom in the high-fidelity model though a projection process.
In general, data fits and lower-fidelity models are widely considered much easier to
implement than ROMs, as they generally do not require any modifications to the
underlying simulation code; however, because ROMs apply a projection process
directly to the equations governing the high-fidelity model, one can often
make much stronger performance guarantees (e.g., of structure preservation, of
accuracy via adaptivity) and perform more accurate \textit{a
posteriori} error analysis (e.g., via \textit{a posteriori} error bounds or
error models) with ROMs. Despite these benefits, the practical
challenges of implementing nonlinear model-reduction techniques in large-scale codes often
precludes their adoption in practice; this occurs because na\"ive implementations
require modifying low-level operations and solvers for each simulation code of interest.
This implementation strategy is simply not practical or sustainable
in many modern settings, because industrial simulation codes often evolve rapidly,
institutions may employ dozens of simulation codes for different analyses,
and commercial codes typically do not expose the required low-level
operators and solvers. This challenge poses perhaps the
single largest barrier to widespread adoption of model
reduction in industrial applications.

To this end, this paper presents an open-source project that aims to
mitigate the implementation burden of nonlinear model reduction
in large-scale applications without compromising performance.
We organize the introduction by first discussing existing packages for ROMs,
some of their limitations, and then describe in detail what we propose.

\subsection*{Overview of the current software landscape}
As previously mentioned, from a software standpoint, most approaches
developed so far for ROMs have relied on intrusive implementations into each code of interest,
see, e.g., the work in~\cite{libMeshPaper:2006, Knezevic:2011, feelPP:2013, 
HesthavenRozzaStamm2015, Stabile2017CAIM, Stabile2017CAF, ITHACADG, ITHACASEM}.
This is impractical for two main reasons: first, each existing method needs
to be re-implemented in any application adapting to various data structure and model;
second, any time a new ROM algorithm is designed,
it would require a separate implementation within each individual code.
To overcome this barrier, recent works such as \modred~\cite{modred:2014},
\librom\footnote{https://github.com/LLNL/libROM}, \pymor~\cite{pymor:2016},
and \pyrom~\cite{pyrom:2019} have sought to provide frameworks for
projection-based ROMs and related capabilities that can be used
by external codes without requiring them to add model-reduction-specific code.

\librom~is a C++ library mainly focused on providing functionalities to compute
proper orthogonal decomposition (POD) basis vectors from a matrix
of samples state vectors. The key feature is that it supports the parallel,
incremental SVD algorithms described in \cite{Brand:2002}.
Since the construction of the reduced order model from the basis vectors
is not a part of \librom~and must be done by the user, we do not further discuss this library.

\modred~\cite{modred:2014} is a Python library containing implementations of Proper Orthogonal
Decomposition (POD) \cite{POD}, balanced POD (BPOD) \cite{willcox2002bmr,rowley2005mrf},
Petrov--Galerkin projection (limited only to linear systems),
and Dynamic Mode Decomposition (DMD) \cite{schmid2010dynamic}.
To use \modred, the user needs to provide (a) a vector object supporting
addition and scalar multiplication;
(b) a function to compute inner products; and (c) a vector handle class with a get and a set method.
The authors of \modred~state in \cite{modred:2014} that the overhead is
small, and if performance becomes a problem, the user is advised to write
required kernels in a compiled language (e.g., C/C++ and Fortran)
and wrap them into Python modules for increased speed.

\pymor~\cite{pymor:2016} is a Python library under active development,
providing reduced-basis (RB) capabilities and aims to be user friendly
and easily integrable with third-party partial-differential-equations (PDE) solvers.
It is similar in concept to \modred, but it provides a more complete and robust
framework. Its design is based on the observation that all
high-dimensional operations in RB methods can be expressed in terms
of a set of elementary linear algebra operations.
\pymor~can be seen ``as a collection of algorithms
operating on \code{VectorArray}, \code{Operator}, and \code{Discretization}
objects'', where the \code{Discretization} object stores the structure of a discrete
problem in terms of its operators.
\pymor~offers two main ways to be used by an application.
One requires the application to output data to disk
(or other persistent storage) from which \pymor~reads/loads what is needed.
The alternative (and preferred) way requires the
application to be re-compiled as a Python extension module,
thus allowing \pymor~to access directly the application's data structures.
\pymor~has valuable capabilities, e.g., its own basic discretization toolkit,
the ability to run the full offline phase.

\pyrom~\cite{pyrom:2019} is a fairly recent Python framework supporting
only Galerkin projection with a basis obtained via POD or DMD.
The work explicitly cites \pymor~\cite{pymor:2016}, but it does
not clearly articulate how it distinguishes itself from \pymor.
The article states that the main operating protocol is one
where \pyrom~reads data from user-supplied files; this approach is
clearly not feasible for large-scale applications, where I/O is extremely expensive.
Further, the work does not articulate how \pyrom~can interface
with application codes written in languages other than Python.

\subsection*{Limitations of current approaches}
We now describe our viewpoint on some weaknesses of the frameworks introduced above.

\paragraph{Limited to continuum systems:\nopunct}
\modred, \pymor~and \pyrom~assume an application to be expressed
as a system of PDEs, and formulate their
ROM methods based on the discrete system stemming from the
spatial discretization of the governing PDEs.
This is applicable to continuum models, but not to naturally
discrete systems such as particle models.
%
%
\paragraph{Limited to spatial reduction:\nopunct}
\modred, \pymor~and \pyrom~support model reduction methods that reduce only
the number of \textit{spatial} degrees of freedoms. They do not support
space--time techniques that enable a reduction of the number of temporal
degrees of freedom
\cite{urban2012new,yano2014space,baumann2016space,Choi:2019}, which is
particularly important for long-time-integration problems.
\paragraph{Limited to linear subspaces:\nopunct}
\modred, \pymor~and \pyrom~focus only on \textit{linear} subspaces, without
discussing potential support for \textit{nonlinear} manifolds.  This raises
questions as to the extensibility of these libraries to promising recent model-reduction
techniques that operate on nonlinear manifolds \cite{gu2008model,deeprom}.
\paragraph{Python as the core development language:\nopunct}
\modred, \pymor~and \pyrom~employ Python as the core development language.
This seems counterintuitive, as---for performance reasons---most Python
packages such as CPython, NumPy and SciPy are natively written in C, and bindings
are provided to expose them to Python.
This approach is also being increasingly adopted by
large-scale numerical libraries,
e.g., PETSc\cite{petsc-user-ref, petsc-efficient}, 
Trilinos\footnote{https://github.com/trilinos}.
Furthermore, the authors of \modred~and \pymor~themselves explicitly advise users
(regardless of the application's language) to write computational kernels
in a compiled language, and create wrappers to use within Python.
We believe, and discuss in more detail later, that a more advantageous
approach would be to use a compiled language as the core, and to create
bindings to expose these functionalities in a interpreted, high-level programming language.
\paragraph{Need for application-specific Python bindings:\nopunct}
Related to the item above, \modred, \pymor~and \pyrom~require application-specific
Python bindings for any application written in a compiled language like C/C++/Fortran
or any other non-Python language.
All three libraries require bindings to wrap the application's data structures.
In addition to this, \pymor~requires bindings to wrap the PDE-discretization operators.
We believe this approach can be problematic for several reasons.
First, the development of these Python bindings may be difficult or even unfeasible.
This is especially true for legacy codes and more modern high performance
computing (HPC)-oriented packages that use complex C++ templates
and structures, like OCCA\cite{medina2014occa}
or Kokkos~\cite{CarterEdwards:2014}\footnote{https://github.com/kokkos}.
This is critical because, for performance reasons, most HPC applications
and libraries targeting large-scale simulations
are written in C, C++, or Fortran using various types of data structures.
Examples include PETSc, Trilinos, Chombo, ScaLAPACK, ALGLIB, Blaze,
OpenFOAM, OCCA, UPC++, mlpack, NAMD, Charm++, LAMMPS, Espresso++, Gromacs, AMReX.
While \pymor~developers offer help to integrate external PDE solvers,
this can remain an obstacle for productivity due to the turnaround time
of this process.
The authors of \pymor~\cite{pymor:2016} aim to overcome this issue by shipping directly
with the library suitable pre-made Python bindings for FEniCS, deal.II, Dune, and NGSolve.
However, these pre-made bindings impose maintainability constraints to \pymor~developers,
as they need to keep them up-to-date with the advancement of the corresponding libraries.
Second, beside the technical challenge, the burden of writing bindings
in a language different than the application's native one could pose
a productivity barrier to HPC developers.
Given the choice, we believe developers would typically favor having access to libraries
written in the language native to their application.
\paragraph{Limited applicability to modern C++ with static polymorphism:\nopunct}
Modern C++ applications and libraries leverage compile-time (or static) polymorphism.
A critical example of its use is for performance portability purposes,
see e.g., OCCA, Kokkos and SYCL.
These programming models are effectively built as C++ extensions, and aimed at
alleviate the user from needing to know architectures-specific details
to write performant code.
They expose a general API allowing users to write code only once, and internally
exploit compile-time polymorphism to decide on the proper data layout, and
instantiate the suitable backend depending on the target threaded system,
e.g.,~OpenMP, CUDA, OpenCL, etc.
Since Python is a dynamic language (types are inferred during interpretation),
static polymorphism does not exist. Consequently, C++ applications
substantially relying on static polymorphism, especially for complex use cases,
would lose all its benefits if they need to interface with Python directly.
\paragraph{Programming model limitations:\nopunct}
\modred, \pymor~and \pyrom~omit any discussion and demonstration of
their interoperability with applications exploiting emerging programming models,
e.g.,~fully distributed task-based models and hybrid MPI+X, where X
represents shared memory parallelism via threads, vectorization,
tasking or parallel loop constructs.
We believe that supporting interoperability with these programming models
is a critical feature, because these have arisen as the leading competitors
to address challenges due to hybrid architectures,
increased parallelism and memory hierarchies, and will become increasingly more important.
%

\subsection*{What we propose}
In this paper, we present \pressio\footnote{https://github.com/Pressio},
a computational framework aimed at enabling projection-based model reduction
for large-scale nonlinear dynamical systems. We highlight here its main features and
how we believe it addresses the aforementioned limitations of
existing frameworks for model reduction.
\paragraph{Applicable to a general ODE systems:\nopunct}
\pressio~provides ROM capabilities that are applicable to any system
expressible as a parameterized system of ordinary differential equations (ODEs) as
\begin{equation}
\label{eq:govern_eq}
\frac{d\state}{dt} = \velocity(\state,t;\params),\quad \state(0;\params) =
\initState(\params),
\end{equation}
where $\state: [0, \tfinal] \times \paramDomain \rightarrow
\mathbb{R}^\fomDof$ denotes the state,
$\initState:\paramDomain\rightarrow\RR{\fomDof}$ denotes the initial state,
$\params\in\paramDomain\subseteq\RR{\nparams}$ denotes the system parameters, $\velocity:\RR{\fomDof}\times
[0,\tfinal]\times\paramDomain\rightarrow \RR{\fomDof}$
with $\velocity:(\dstate,\dtime,\dparams)\mapsto\velocity(\dstate,\dtime,\dparams)$
denotes the velocity that may
be linear or nonlinear in its first argument,
and $t\in[0,\tfinal]$ denotes time with $\tfinal>0$ denoting the final time.
We note that the model form \eqref{eq:govern_eq} is highly expressive, as it
may be derived from the spatial
discretization of a PDE problem or from naturally discrete systems (e.g.,
molecular-dynamics problems).

\paragraph{Minimal API:\nopunct}
An application needs to satisfy a minimal application programming interface (API).
This consists of exposing, for a given state $\state$, time $t$,
and parameters $\params$, the velocity vector $\velocity(\state,t;\params)$
and the action of the Jacobian matrix
$\partial \velocity(\state,t;\params)/\partial\dstate$ on a dense matrix.

\paragraph{Methods:\nopunct} \pressio~supports model-reduction methods applicable to
both steady and unsteady problems. For the latter, in particular,
it provides algorithms to reduce both the spatial and temporal degrees of freedom.
A detailed discussion of the supported methods is provided in \S~\ref{sec:romtheory}.

\paragraph{Trial manifold abstraction:\nopunct} We exploit software design and
abstractions to generalize the concept of the trial manifold onto which
the full-order model (FOM) system is projected, yielding support for both linear trial manifolds (i.e.,
linear trial subspaces) and nonlinear trial manifolds.

\paragraph{Suitable for complex nonlinear problems:\nopunct}
\pressio~provides ROM capabilities that are meant for complex nonlinear problems,
including the least-squares Petrov--Galerkin (LSPG) projection technique
\cite{Carlberg:2011,CarlbergGalVsLSPG}.
This technique exhibits advantages over the popular Galerkin projection
because it (a) guarantees optimality of the time-discrete ROM,
(b) yields smaller discrete-time \textit{a posteriori} error bounds, (c) and often
yields improved accuracy.
These properties have been demonstrated for unsteady
CFD applications \cite{CarlbergGnat,CarlbergGalVsLSPG}.
Furthermore, the LSPG method has been applied successfully to highly
nonlinear problems including transonic and hypersonic CFD cases \cite{washabuagh2016steadyROM, Blonigan2020}.

\paragraph{C++ core:\nopunct}
\pressio~is built on a C++(11) core, following a model consistent with the
one used by CPython, NumPy, SciPy, as well as other HPC libraries like
Trilinos and PETSc.
This approach enables us to have direct integration with native C/C++ applications,
and to build multiple front ends of various languages all relying on a single,
compiled, optimized, and robust backend.
One could argue that Fortran could have been another choice, which would bring us
into a historical debate between Fortran and C/C++.
We opted for C++ because (a) a substantial subset of the large-scale codes
across the United States national laboratories is implemented in C++, and (b) C++ seems
better suited than Fortran to design complex data structures and abstractions.
%

\paragraph{No bindings for C++ applications:\nopunct} Any C++ application can
use \pressio~simply as a regular library, thus avoiding any bindings.
This is different than \modred, \pymor~and \pyrom, which require any C++
application to develop individual Python bindings.
Eliminating the need for bindings benefits productivity which, in turn,
maximizes the impact and appeal of \pressio~not only across the national
laboratories,
but also on any other scientific code written in C++.
We remark that this does not mean that any C++ application ``magically''
works with \pressio: as stated previously, an application must satisfy
a minimal but specific API, which is discussed in
\S~\ref{ssec:pressio_api}.
If an application does not natively satisfy the API, its developers/users would
need to expose it by working \textit{within} the application
itself using the \textit{same} programming language.
Hence, in any scenario, users and application developers operate within their domain of expertise,
without needing to resort to a different language to interface to \pressio.
%

\paragraph{Compile-time benefits and easy deployment:\nopunct}
\pressio~is a header-only library written in C++(11), allowing us
to fully exploit metaprogramming and perform extensive checks at \textit{compile time}.
Being header only, it does not need to be compiled, so the user can simply
point to the source code, yielding a straightforward deployment.

\paragraph{Seamless interoperability with arbitrary programming models:\nopunct}
\pressio~is designed without making any assumption on the programming model
or data types used by an application. Hence, it seamlessly supports arbitrary
data types and arbitrarily complex programming models,
e.g., MPI+X with X being a graphics processing unit (GPU), OpenMP or local tasking.
%


\paragraph{No binding code for Python applications:\nopunct}
We leverage the modern, C++11 library \pybind\footnote{https://github.com/pybind/pybind11}
to create Python bindings to \pressio. This enables any user having an
application fully written in Python to use the C++ backend of \pressio~to run ROMs.
No user-defined bindings are needed (see \S~\ref{ssec:pressio_pythoninterface} for a detailed discussion).
This is a key result for a two reasons. First, to support this feature, we only
had to develop Python bindings in the core library \textit{once}, which were
then immediately usable by \textit{any} Python application. Second, this
exempts Python users from having to write any binding to interface with
\pressio.

\subsection*{What are the drawbacks of \pressio?\nopunct}
First, being written in C++, it is not directly usable in Fortran applications.
However, as mentioned above, this can be
addressed by writing Fortran bindings, as it has been done for other libraries,
e.g.,~hypre\footnote{https://github.com/hypre-space/hypre}
(which uses Babel\footnote{https://computing.llnl.gov/projects/babel-high-performance-language-interoperability} to write its Fortran interface),
and ForTrilinos\footnote{https://github.com/trilinos/ForTrilinos}
(which is based on SWIG\footnote{http://www.swig.org/}).

Second, the use of metaprogramming might constitute a barrier to entry
for developers who are not familiar with this technique.
However, considering that generic programming is becoming increasingly
more important and widespread, it is a useful skill to learn. Hence, working
on \pressio~can be advantageous for developers and, at the same time,
beneficial to the actual code development.

Third, by relying on a minimal, ODE-oriented API, when a new model-reduction
technique is developed, one needs to ensure (if possible) that its formulation
is compatible with the same API. This might pose some limitations, e.g., for
certain structure-preserving methods that leverage the underlying second-order
or Hamiltonian structure \cite{chaturantabut2016structure} or the underlying
finite-element \cite{farhat2015structure}, finite-volume
\cite{carlberg2018conservative}, or discontinuous Galerkin
\cite{yano2019discontinuous} spatial discretization.  In the worst case
scenario, this can be addressed by augmenting the API for \pressio~for new
developments; indeed, we have already provided the required extension to
enable conservative LSPG \cite{carlberg2015preserving} for finite-volume
models.

\subsection*{Paper organization}
The paper is organized as follows. In \S~\ref{sec:romtheory}, we
describe the formulation of the ROMs currently supported
in \pressio, as well as those planned for the next release.
In \S~\ref{sec:pressio}, we discuss the implementation,
the application programming interface (\S~\ref{ssec:pressio_api}),
support for arbitrary data types (\S~\ref{ssec:pressio_arbitrarydata}),
and the Python bindings (\S~\ref{ssec:pressio_pythoninterface}).
Representative results are discussed in \S~\ref{sec:results}, and
conclusions in \S~\ref{sec:conclusions}.

\section{Reduced-Order Models Formulation}
\label{sec:romtheory}

This section presents the mathematical formulation of the
ROMs currently supported in \pressio, as well as some details
about those scheduled for integration in upcoming releases.

We consider the original computational model---which we refer to as the
full-order model (FOM)---to be a dynamical system expressible as a system of
parameterized ODEs of the form \eqref{eq:govern_eq}, which can be expressed
equivalently as
\begin{equation}
\label{eq:govern_eqRes}
\res\left(\difftime{\state}, \state, t;\params\right) = \zero, \qquad \state(0;\params) = \initState(\params),
\end{equation}
where $\res:\RR{\fomDof}\times\RR{\fomDof}\times[0,\tfinal]\times\paramDomain\rightarrow\RR{\fomDof}$
denotes the (time-continuous) residual function
defined as $\res \defeq \difftime{\state} - \velocity(\state,t;\params)$.

\subsection{Trial manifold}
Projection-based model-reduction methods seek an approximate solution
$\aprxState (\approx \state)$ of the form
\begin{equation}\label{eq:state_approx}
\aprxState(t; \params) = \stateRef(\params) + \romToFomMapping(\romState(t;\params)),
\end{equation}
where $\aprxState:[0,\tfinal]\times\paramDomain\rightarrow \stateRef(\params)+
\manifold\subseteq\RR{\fomDof}$ and
$\manifold \defeq \{ \romToFomMapping(\gRomState)\,|\, \gRomState\in\RR{\dofRom}\}\subseteq\RR{\fomDof}$
denotes the (linear or nonlinear) trial manifold of $\RR{\fomDof}$.
Here, $\stateRef:\paramDomain\rightarrow\RR{\fomDof}$ denotes a parameterized
reference state and $\romToFomMapping:\drdstate \mapsto \romToFomMapping(\drdstate)$
with $\romToFomMapping: \mathbb{R}^{\dofRom} \rightarrow \mathbb{R}^{\fomDof}$ and
$\dofRom \leq \fomDof$ denotes the parameterization function mapping low-dimensional
generalized coordinates $\romState$ to the high-dimensional state approximation $\aprxState$.
Henceforth, we will refer to this mapping as the ``decoder'' function,
consistently with terminology employed in \cite{deeprom}, where this function
associated with a convolutional decoder.
The approximate velocity can be obtained via chain rule yielding
\begin{equation}\label{eq:velo_approx}
\difftime{\aprxState}(t; \params) = \jacOfRomFomMapping(\romState(t;\params))
\difftime{\romState}(t;\params),
\end{equation}
where $\jacOfRomFomMapping:\gRomState\mapsto \frac{d\romToFomMapping}{d \drdstate}(\gRomState)$ with
$\jacOfRomFomMapping:\RR{\dofRom}\rightarrow\mathbb{R}^{\fomDof \times \dofRom}$
denoting the Jacobian of the decoder.

\paragraph{{Linear trial manifold}}
When $\romToFomMapping$ is a linear mapping, one recovers the classical
affine trial subspace. In this case, the decoder can be expressed as
$\romToFomMapping:\drdstate \mapsto \romBasis\drdstate$ for some trial basis matrix $\romBasis \in
\RRstar{\fomDof \times \dofRom}$, where
$\RRstar{n\times m}$ denotes the set of full-column-rank $n\times m$ matrices.
In this case, the trial manifold is affine with
$\manifold = \range(\romBasis)$ and the Jacobian a constant matrix
$\jacOfRomFomMapping(\romState(t;\params)) = \jacOfRomFomMapping= \romBasis$,
leading to the following expression of the approximate state and velocity
\begin{equation}
\aprxState(t; \params) = \initState(\params) +
\romBasis \romState(t;\params) \quad \text{and} \quad
\difftime{\aprxState}(t; \params) = \romBasis \difftime{\romState}(t;\params).
\end{equation}

\subsection{Manifold Galerkin ROM}
\label{ssec:romtheory_galerkin}

\manifGalCap~projection as proposed in \cite{deeprom} can be derived by
minimizing the time-continuous residual over the trial manifold.
The resulting model can be obtained by substituting the approximate state
\eqref{eq:state_approx} and corresponding velocity
\eqref{eq:velo_approx} into Eq.~\eqref{eq:govern_eqRes},
and minimizing the (weighted) $\ell^2$-norm of the resulting residual, which yields
the ODE system
\begin{equation}\label{eq:cont_opt_problem}
	\difftime{\romState}(t;\params)  =
	\underset{\gVelo\in\RR{\dofRom}}{\arg\min} \left
	\| \weightMat \res\left(\jacOfRomFomMapping(\romState(t;\params)) \gVelo,
        \stateRef(\params) + \romToFomMapping(\romState(t;\params)), t;\params \right)
	\right\|_2^2,
\end{equation}
which can be written equivalently as
\begin{equation} \label{eq:cont_rom}
	\difftime{\romState}(t;\params) =
        ( \weightMat \jacOfRomFomMapping(\romState(t;\params))\pseudoInv \weightMat
	\velocity\left(\stateRef(\params)+\romToFomMapping(\romState(t;\params)), t; \params
	\right),
\end{equation}
where the superscript $\pseudoInvSymb$ denotes the Moore--Penrose pseudoinverse,
$\romState(0;\params)=\initRomState(\params)$ is the reduced initial
condition (e.g., computed by performing orthogonal projection of the
$\initState(\params)$ onto the trial manifold \cite{deeprom}),
and $\weightMat \in \RR{z\times\fomDof}$ with $z \leq \fomDof$
is a weighting matrix that enables the definition of a weighted (semi)norm.
The ODE system \eqref{eq:cont_rom} can be integrated in time using any time stepping scheme.

If we select the $\fomDof \times \fomDof$ identity matrix,
i.e.~$\weightMat = \identity$, the projection is
typically insufficient to yield computational savings when the velocity is
nonlinear in its first argument, as
the $\fomDof$-dimensional nonlinear operators
must be repeatedly computed, thus precluding an $\fomDof$-independent online
operation count.
To overcome this computational bottleneck, one can choose
$\weightMat$ to have a small number of nonzero columns. When the residual
Jacobian is sparse, this  sparsity pattern
leads to ``hyper-reduction'', which yields an $\fomDof$-independent online
operation count by approximating the
nonlinear operators by computing a small ($\fomDof$-independent) subset of their
elements.
Examples include collocation, i.e., $\weightMat = \collocMat$ with
$\collocMat$ comprising selected rows of the identity matrix,
and the GNAT method  \cite{Carlberg:2011, CarlbergGnat}, i.e., $\weightMat =
\gnatWMat$ with $\phires\in\RRstar{\fomDof\times z}$ comprising a basis matrix for the residual.


\manifGalCap~is currently supported in \pressio~for both linear and nonlinear
trial manifolds. For the linear case, the user needs to provide the constant
basis matrix $\romBasis$.
For the nonlinear case, the user is responsible for implementing the decoder
mapping $\romToFomMapping$, which \pressio~uses without knowledge of its implementation details.
In a future release, we plan to add support for a set of default choices
for nonlinear mappings, e.g., convolutional autoencoders as in \cite{deeprom}.
For the weighting matrix, we currently support $\weightMat = \identity$,
the collocation-based hyper-reduction $\weightMat = \collocMat$, and
an arbitrary weighting matrix specified by the user, while the GNAT-based
weighting matrix $\weightMat = \gnatWMat$ is under development.

\subsection{Manifold Least-squares Petrov--Galerkin (LSPG) ROM}
\label{ssec:romtheory_lspg}
In contrast to Galerkin projection, LSPG projection corresponds to minimizing
	the (weighted) $\ell^2$-norm of the \textit{time-discrete} residual over the trial manifold.
Hence, the starting point for this approach the residual ODE formulation
	\eqref{eq:govern_eqRes} discretized in time with an arbitrary
	time-discretization method.
Here, we focus on \textit{implicit} linear multistep methods
for two primary reasons: first, Galerkin and LSPG are equivalent
for explicit schemes~\cite{CarlbergGalVsLSPG};
second, at the time of this writing, implicit linear multistep
schemes are the main implicit schemes supported in \pressio.
Other families of time stepping schemes, e.g., diagonally implicit
	Runge--Kutta, will be supported in future developments;
	\cite{CarlbergGalVsLSPG} describes LSPG for Runge--Kutta schemes in detail.

A linear $k$-step method applied to numerically solve the problem
\eqref{eq:govern_eq} leads to solving a sequence of systems of algebraic equations
\begin{equation}
\label{eq:fomODeltaE}
\res^{n}(\state^{n};\params) = \zero,\quad n=1,\ldots,\nseq,
\end{equation}
where  $\nseq$ denotes the total number of instances,
$\state^{n}$ denotes the state at the $n$th time instance, and
$\res^{n}:\RR{\fomDof}\times\paramDomain\rightarrow\RR{\fomDof}$ denotes
the \textit{time-discrete} residual.
At the $n$-th time instance, we can then write:
\begin{align}
\alpha_0 \state^{n} - \Delta t \beta_0 \velocity(\state^n,t^n;\params)
\sum_{j=1}^{k} \alpha_j \state^{n-j}
\Delta t \sum_{j=1}^{k} \beta_j \velocity(\state^{n-j}, t^{n-j}; \params) = 0,
\end{align}
where $\Delta t \in \mathbb{R}_{+}$ denotes the time step,
$\state^{k}$ denotes the numerical approximation to $\state(k \Delta t; \params)$, the
coefficients $\alpha_j$ and $\beta_j$, $j=0,\ldots, k$ with
$\sum_{j=0}^k\alpha_j=0$ define a particular multistep scheme,
and $\beta_0\neq 0$ is required for implicit schemes.
For simplicity, we assume a uniform time step $\Delta t$ and a fixed number
of steps $k$ for each time instance, but extensions to nonuniform time grids
and a non-constant value of $k$ are straightforward.
For example, setting $k=1$ and $\alpha_0=\beta_0=1$, $\alpha_1=-1$, $\beta_1=0$,
yields the backward Euler method.

LSPG projection is derived by substituting the approximate state
\eqref{eq:state_approx} in the time-discrete residual \eqref{eq:fomODeltaE}
and minimizing its (weighted) $\ell^2$-norm, which yields a sequence of
residual-minimization problems
\begin{equation}\label{eq:disc_opt_problem}
	\romState^n(\params)  =
	\underset{\gRomState\in\RR{\dofRom}}{\arg\min}
        \left\| \weightMat \res^{n}\left(
	\stateRef(\params)+\romToFomMapping(\gRomState);\params)\right)\right\|_2^2,\quad
	n=1,\ldots,\nseq
\end{equation}
with
initial condition $\romState(0;\params)=\initRomState(\params)$. As in
\S~\ref{ssec:romtheory_galerkin},
$\weightMat \in \RR{z\times\fomDof}$ with $z \leq \fomDof$
denotes a weighting matrix that can enable hyper-reduction.

\paragraph{LSPG formulation for stationary problems}
\manifLSPGCap~can be applied to steady-state problems by computing
$\state(\params)$ as the solution to Eq.~\eqref{eq:govern_eq} with $\difftime{\state}=\zero$,
which yields
\begin{equation}\label{eq:govern_eq_steady}
\velocity(\state;\params)=\zero,
\end{equation}
where the time-independent velocity is equivalent to the residual of the stationary-problem.
Substituting the approximate state
\eqref{eq:state_approx}
into Eq.~\eqref{eq:govern_eq_steady} and minimizing the $\ell^2$-norm
of the residual yields
\begin{equation}\label{eq:disc_opt_problem_steady}
\romState(\params)  =
\underset{\gRomState\in\RR{\dofRom}}{\arg\min}
\left\|
\weightMat\velocity\left(
\stateRef(\params)+\romToFomMapping(\gRomState);\params)\right)\right\|_2^2,
\end{equation}
which has the same nonlinear-least-squares form obtained for a time-dependent
problem shown in Eq.~\eqref{eq:disc_opt_problem}.

\paragraph{Constrained LSPG formulation}
If the minimization problems \eqref{eq:disc_opt_problem} and
\eqref{eq:disc_opt_problem_steady} are equipped with constraints,
one obtains unsteady and steady constrained LSPG problems, respectively.
An example within this category of models is the conservative LSPG formulation
in \cite{carlberg2018conservative}, which
ensures the reduced-order model is conservative over subdomains of the
problem in the case of finite-volume discretizations. This method relies on a
nonlinear equality constraint to enforce conservation.  The reader is referred
to \cite{carlberg2018conservative} for additional details on conservative LSPG
projection, including feasibility conditions of the associated optimization
problems, and error bounds.

\pressio~supports \manifLSPG~and conservative LSPG~for both
steady and unsteady problems. As with manifold Galerkin projection, it
supports weighting matrices $\weightMat = \identity$,
the collocation-based hyper-reduction approach $\weightMat = \collocMat$,
or an arbitrary weighting matrix specified by the user.
Native support for the GNAT-based weighting matrix is under development.

\subsection{Space--time model reduction}
\label{ssec:romtheory_others}
In \S~\ref{ssec:romtheory_galerkin} and \S~\ref{ssec:romtheory_lspg}, we presented ROMs
that reduce the \textit{spatial} dimensionality of a dynamical system; this yields
a ROM that still needs to be resolved in time via numerical integration.
In some cases, the reduced system can be integrated with a time step
larger than the one used for the full-order model, see~\cite{Bach:2018}.
However, this still imposes a limitation to the achievable computational savings.
To address the time dimension directly, one can employ space--time
model-reduction methods that simultaneously
reduce both the number spatial and temporal degrees of freedom of the ODE system.
To address this need, we are currently implementing in \pressio~the
space--time least-squares Petrov--Galerkin (ST-LSPG)
projection method \cite{Choi:2019} and the windowed least-squares (WLS) method \cite{ParishWLS:2019}.
Because these techniques will be released in the future, we briefly describe here
the main idea behind them, but omit the full mathematical details.
For a complete formulation of ST-LSPG and WLS, the reader is
directed to the aforementioned references.

ST-LSPG relies on a low-dimensional space--time trial manifold, which---in the
case of a linear space--time trial manifold---can
be obtained by computing tensor decompositions of snapshot matrices constructed
from spatio-temporal state data. The method then computes discrete-optimal
approximations in this space--time trial subspace by
minimizing the (weighted) $\ell^2$-norm of the space--time residual, which
corresponds to the concatenation of the time-discrete residual over all time steps.

The windowed least-squares (WLS) \cite{ParishWLS:2019} method is a model-reduction
approach comprising a generalization of Galerkin, LSPG, and ST-LSPG.
WLS operates by sequentially minimizing the dynamical system residual within
a low-dimensional trial manifold over time windows. WLS can be formulated for two types
of trial manifolds: \textit{spatial trial manifolds} that reduce the spatial
dimension of the full-order model, and \textit{space--time trial manifolds}
that reduce the spatio-temporal dimensions of the full-order model.

\section{\pressio}
\label{sec:pressio}

The core component of \pressio~is a C++(11) header-only library that leverages
generic programming (i.e.,~templates) to enable support for native C++
applications with arbitrary data types.
Metaprogramming is used for type checking and introspection,
to ensure that types are compatible with the design and requirements.
Being header-only, \pressio~does not need to be separately compiled,
packaged, or installed; the user must only point to the source code.
The development is based on the following objectives:
(a) support for general computational models (data structures, problem types),
(b) heterogeneous hardware (many-core, multi-core, accelerators),
(c) performance, and (d) extensibility.
The key advantage of having a single framework is that
when a new method is developed, it can be immediately
used by any application already using \pressio.
\pressio~is modularized into separate packages as follows:
\bit
\item \code{mpl}: metaprogramming functionalities
\item \code{utils}: common functionalities, e.g., I/O helpers, static constants, etc.
\item \code{containers}: data strutures 
\item \code{ops}: linear algebra operations
\item \code{qr}: QR factorization functionalities
\item \code{svd}: singular value decomposition (SVD) functionalities
\item \code{solvers}: linear and nonlinear solvers
\item \code{ode}: explicit and implict time steppers and integrators
\item \code{rom}: reduced-order modeling methods
\eit
The order used above is informative of the dependency structure.
For example, every package depends on \code{mpl},
but \code{qr} depends only on \code{mpl, utils, containers}.
The \code{rom} package resides at the bottom of the hierarchy and depends on the others.
Each package contains corresponding unit and regression tests.

\subsection{What API does \pressio~require from an external application?\nopunct}
\label{ssec:pressio_api}

The design of \pressio/\code{rom} is based on the observation
that all methods listed in \S~\ref{sec:romtheory} can be
applied to any dynamical system of the general form~(\ref{eq:govern_eqRes}),
and only require access to the velocity, $\velocity$, and the \textit{action}
of the Jacobian matrix $\fomJac$, on an operand, $\Boper$,
which is a tall, skinny matrix
(e.g., the decoder Jacobian
$d\romToFomMapping/d\drdstate$).
Querying the action of the Jacobian on an operand rather than the Jacobian itself is
essential, because in many applications the Jacobian itself
cannot easily be explicitly assembled or stored.

We envision two scenarios: one in which an application natively
supports the above functionalities, and one in which it does not.
In the first scenario, minimal work (e.g.,~rearranging some functions in the native code)
is needed to adapt the application to use \pressio/\code{rom}.
We believe this best case scenario to be a reasonable expectation
from applications using expressive syntax, well-defined abstractions and encapsulation.
These software engineering practices guide developers to design codes
using proper abstractions directly expressing the target mathematical concept.
For systems of the form (\ref{eq:govern_eqRes}), this means implementing the
velocity $\velocity$ as a self-contained
free function or as a class method encapsulating the implementation details.
This kind of abstraction and design would satisfy the best practices for
software engineering and also naturally fits our expressive API model.
To further emphasize this point, we remark that the same expressive model
is being used/expected by other well-established libraries,
e.g.,~Sundials\cite{hindmarsh2005sundials}, SciPy.ode. If an application needs to
be interfaced with these libraries, the developers would need to meet
a similar, expressive, intuitive API designed for dynamical systems.

When an application does not readily expose the required functionalities,
the user or developer needs to expose them. This can be done, e.g.,
by writing an adapter class inside the application.
Even for this scenario, we argue that the effort needed is
easily manageable, and even useful, for the following reasons:
(a) the adapter class needs to be developed using the same
language of the application; (b) to write the adapter,
a developer/user would still operate within their domain of knowledge;
and (c) once this adapter class is complete, the application
would gain (almost) seamless compatibility
with other libraries targeting dynamical systems
since most of them are based on the same API model.

\begin{figure}
  \tikzstyle{block} = [draw, fill=green!20, rectangle, minimum height=3em]
  \tikzstyle{fbox} = [rectangle, draw, line width=0.5mm, densely dashed, inner sep=2mm]
  \centering
  \resizebox{0.525\columnwidth}{!}{ \input{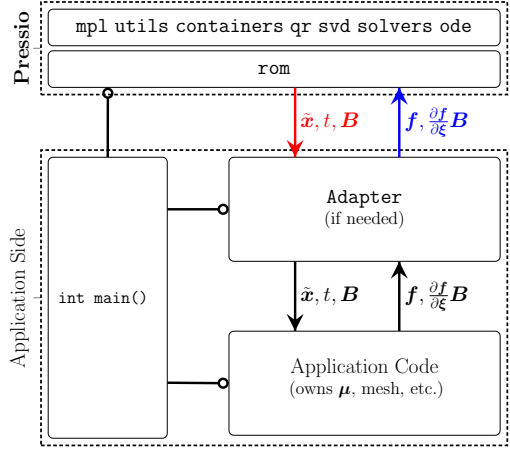} }
  \caption{Schematic of the interaction between \pressio~and the application,
  highlighting elements belonging to \pressio~and those belonging to the application.
  The solid, black line ($-\circ$) indicates that the main function
  is where all the various components are used.
  The red arrow is used to indicate information/data computed by and exiting \pressio,
  while blue is used to denote information/data computed by the application
  and entering \pressio. For a steady problem, the schematic would
  be similar without the time dependence.}
\label{fig:pressio}
\end{figure}

Figure~\ref{fig:pressio} shows a schematic of the interaction
between \pressio/\code{rom} and a generic application.
The application owns the information defining the problem,
i.e., the parameters $\params$, mesh, inputs, etc.. \pressio~is used as an external library,
and the interaction can be seen simply as an exchange of
information: for a given approximate state $\aprxState$, time $t$, and operand $\Boper$,
\pressio/\code{rom} queries the application for the associated velocity
$\velocity(\aprxState,t;\params)$ and Jacobian action $\aprxFomJac \Boper$.
In the schematic, a red arrow indicates the information/data exiting
\pressio~and entering the application, while a blue arrow denotes the reverse.
If the adapter is not needed, then \pressio~interacts directly with the core application classes.
When the problem is steady, time dependency simply drops.

Listing~\ref{lst:adapter} shows an example of a valid C++ adapter class
satisfying the API for Galerkin and LSPG for unsteady problems.
The listing shows four type aliasing declarations and four methods.
If any of them are missing, or the syntax shown is not met exactly, a compile-time error is thrown.
This static checking allows us to ensure the correctness of the API.
For performance reasons, the non-void methods are only called once
during the setup stage to copy construct the objects.
More details on the overall software design are presented in the supplementary material.

\lstset{style=cppstyle}
\begin{lstlisting}[
    float,
    caption={Skeleton C++ adapter class satisfying the unsteady API for Galerkin and LSPG ROMs.},
    label={lst:adapter}]
class SampleAdapterClass{
  //...
public:
  /* C++11 type aliasing declarations that Pressio detects */
  /* this is equivalent to doing: typedef ... scalar_type  */
  using scalar_type       = /*application's scalar type       */;
  using state_type        = /*              state type        */;
  using velocity_type     = /*              velocity type     */;
  using dense_matrix_type = /*              dense matrix type */;
  //...

  // compute velocity, f(x,t;...), for a given state, x(t)
  void velocity(const state_type  & x,
                const scalar_type & t,
                velocity_type     & f) const;

  // given current state x(t):
  // 1. compute the spatial Jacobian, df/dx
  // 2. compute A=df/dx*B, B is typically a skinny dense matrix
  void applyJacobian(const state_type        & x,
                     const dense_matrix_type & B,
                     const scalar_type       & t,
                     dense_matrix_type       & A) const;

  // overload called once to construct an initial object
  velocity_type velocity(const state_type  & x,
                         const scalar_type & t) const;

  // overload called once to construct an initial object
  dense_matrix_type applyJacobian(const state_type        & x,
                                  const dense_matrix_type & B,
                                  const scalar_type & t) const;
\end{lstlisting}
All the methods are \code{const}-qualified to ensure const-correctness.
\pressio~does not cast away the const qualification of
the adapter/application object to preserve its visibility.
The user can choose to implement the adapter class' methods such that no
modification happens for the data members, or explicitly use the \code{mutable} C++ keyword.
The \code{mutable} keyword allows one to qualify data members that can
be modified within those \code{const}-qualified methods.
This is typically used for cases, like this one, where one may have to mutate
the state for an implementation detail, while keeping the visible state of an object the same.

The adapter class' methods accept arguments by reference to avoid
(potentially) expensive copies of large objects.
If an application relies on data structures with ``view semantics'',
see e.g., Kokkos or Tpetra in Trilinos, the user can simply
omit the references in the arguments of the adapter class methods.
View semantics means that an assignment operator performs a shallow-copy operation,
i.e.,~only the memory reference and dimensions are copied. View semantics thus behave
like the C++ shared pointer semantics. While it would still work to accept by reference
objects of types with view semantics, it would be more appropriate to pass them
by copy to ensure proper reference counting.

\subsection{How does \pressio~handle arbitrary data structures?\nopunct}
\label{ssec:pressio_arbitrarydata}

Support for arbitrary data types is enabled via generic programming and type introspection.
Generic programming (i.e.,~C++ templates) implies that, implementation-wise,
\pressio~can naturally accomodate arbitrary types.
Quoting from \cite{Musser:1988}: ``Generic programming centers around the
idea of abstracting from concrete, efficient algorithms to obtain generic
algorithms that can be combined with different data representations to
produce a wide variety of useful software''.
Broadly speaking, we can say that \pressio~contains algorithms
and operations written abstractly as high-level steps to operate on a generic type.
This, however, is a necessary but not sufficient condition to enable arbitrary types.
It is clear, in fact, that some class or functions templates
performing specific operations are applicable only to types satisfying certain conditions.
This is where type introspection comes into play.

Type introspection allows the examination of a type to analyze its properties
at compile-time using metaprogramming techniques.
This is the foundation of design by introspection, a programming paradigm that relies on
performing introspection on components and making decisions during compilation.
For the sake of argument, think of a function template
applicable only to types that can be default constructed.
Introspection can be used to examine the template argument
to detect if it is default constructible, and differentiate code
according to whether this property is satisfied or not.
Another example is the ability to selectively enable/disable functions
or classes based on whether the corresponding template
arguments satisfy some requirements.

The above approach is not exclusive to achieve support for arbitrary types.
One possible alternative is based on inheritance and dynamic polymorphism.
In this case, the library would contain specific base (or abstract)
classes to represent well-defined interfaces for target functionalities.
These abstract classes are used via dynamic polymorphism
inside the algorithms. Users would be able to employ their data types
by inheriting from the base classes, thus creating new types that
are fully compatible with the library algorithms.
This is a valid approach, but has some disadvantages:
(a) templates and type introspection is an approach more
suitable than inheritance to obtain type safety;
(b) dynamic polymorphism has an overhead (due to a virtual table lookup)
for each function invocation; (c) type introspection
fully exploits the power of templates, which is a key
feature of C++; (d) design by introspection
will become even more natural to adopt and use
when C++ concepts become part of the standard.

\pressio~performs introspection on the types instantiated
by the user and, therefore, can decide at compile-time to
select specialized implementations of type-specific algebra operations.
To this end, we have implemented in \pressio~a suite of pre-defined
implementations of basic algebra operations for several widely-used
libraries, e.g., Trilinos, Kokkos, Eigen, Armadillo.
Additional implementations for other packages, like PETSc and hypre, are planned for upcoming releases.
Whenever possible, \pressio~exploits the implementations native
to the target libraries to perform all operations.
This approach allows us to fully exploit the native performance of
these libraries benefiting from the substantial amount of work
done to achieve optimal performance in these libraries.
If an external application uses a custom set of data structures
and operations, or if types-specific operations are not already
supported in \pressio, the user needs to provide a class implementing
specific operations needed to carry out all of the necessary ROM computations,
e.g., dot products, matrix-vector and matrix-matrix products.
One advantage of this approach is that, if needed, the user can ``plug-and-play''
and test various implementations of the low-level algebra operations.

While this approach might resemble the one in \modred, \pymor~and \pyrom,
we highlight the following differences. \modred, \pymor~and \pyrom~require
C/C++/Fortran users to write custom Python---which is not the native
application's language---bindings to wrap their application's data structures.
In addition, \pymor~needs bindings to wrap the application's PDE operators.
\pressio, on the contrary, does not need bindings for the application's
data structures, since these just become template arguments.
Also, \pressio~only requires (in certain cases) a single class containing
a small set of basic algebra operations. We stress that this class with
algebra operations requires little effort from the user,
because (a) it is written in the application's native language;
and (b) it simply involves forwarding calls to operations
and methods already implemented by the application's
native data structures. Finally, \pressio~does not need any information
about the application's spatial operators or discretization.
This stems from the fact that the ROMs in \pressio~are only based
on the assumption that a target system is represented
as a system of ODEs.

\subsection{How does \pressio~support hyper-reduction?\nopunct}
\label{ssec:pressio_hyperreduction}

The notion of hyper-reduction (or sample mesh) was introduced in \S~\ref{sec:romtheory},
and consists of approximating the nonlinear operators (residual and, if needed, the Jacobian)
of the full order model by sampling only a subset of elements.
Equipping a ROM with this technique yields large computational savings, see
e.g., \cite{CarlbergGnat},
and is thus a critical component. A detailed description and comparison of
various hyper-reduction techniques can be found in \cite{CarlbergGnat}.

In \pressio, one can explore the effects of hyper-reduction in two ways.
The first, which we refer to as ``algebraic hyper-reduction'', requires
the user to simply provide the desired weighting matrix, $\weightMat$,
which \pressio~applies to the FOM operators to extract the target subset of elements.
The immediate benefit is a much smaller nonlinear minimization problem to solve,
see Eqs.~\eqref{eq:cont_opt_problem}, \eqref{eq:disc_opt_problem} and \eqref{eq:disc_opt_problem_steady}.
While easy to implement, the main drawback is that the FOM still needs to
compute the full operators, whose cost typically dominates the
cost to solve the minimization problem.
This implies that one cannot expect to obtain substantial savings.
Hence, the algebraic hyper-reduction allows one to just \textit{mimic} the effect of a real implementation.

The second (preferred) way requires an application to have
the capability of evaluating the nonlinear operators
\textit{only} at target locations of the mesh, i.e., a ``sample mesh''.
This scenario considerably reduces the cost of evaluating the FOM's
operators, thus accentuating the benefits of hyper-reduction,
see e.g., \cite{Blonigan2020}.
This second approach, however, is strictly dependent on the
application's numerical method and mesh, and thus needs to be
implemented by the user directly within the application.
We remark, however, that while this task can be intrusive, it is paradoxically
easier to implement for complex applications, e.g.,~those based on an unstructured mesh.
On the one hand, the complexity of these meshes is a disadvantage because
one has to specify the full connectivity between the elements.
On the other hand, the complexity of an unstructured mesh is an advantage because it allows more freedom in
how the elements are assembled, which clearly benefits the construction of the sample mesh.
This will be further discussed in \S~\ref{ssec:results_sparc} in relation
to one of our test cases.

\subsection{\pressioFpy}
\label{ssec:pressio_pythoninterface}

To increase the impact and widen the range of potential users of \pressio,
we have developed \pressioFpy\footnote{https://github.com/Pressio/pressio4py},
a library of Python bindings to \pressio.
\pressioFpy~is implemented using \pybind~\cite{pybind11}, a lightweight
open-source library that leverages C++11 and metaprogramming
to facilitate the interoperability between C++ and Python.
The objective of \pressioFpy~is twofold.
First, to allow \textit{any} Python application based on NumPy to use the
capabilities in \pressio~without any additional glue code.
The only requirement is that the Python application
meets an API similar to the one shown in listing \ref{lst:adapter}.
Second, to enable a Python application to obtain a performance
comparable to the one obtained with a C++ application natively interfacing with \pressio.

Developing \pressioFpy~has required minimal effort, mostly in the form of add-on
functionalities needing minor modifications to the core C++ \pressio~code.
This is a result of two main factors, namely the quality and ease of use of \pybind,
and the modularity of \pressio, which allowed us to develop extensible
and maintainable bindings relatively quickly and easily.
More specifically, the work involved two main stages.
One aimed at enabling \pressio~to support \pybindarr.
The latter is a class template in \pybind~to bind a NumPy array from the C++ side.
Using \pybindarr~allows us to optimize interoperability with Python
since no copy needs to occur when a NumPy array from the Python side is
passed to a C++ function accepting a \pybindarr.
This also allows us to bypass the Python interpreter,
and perform all operations by working directly on the underlying data,
thus yielding a negligible overhead.
A key property of \pybindarr~is that it has view semantics and, therefore,
special attention is needed depending on whether a deep or shallow copy is needed.
To achieve this, we leverage type introspection allowing \pressio~to support
\pybindarr~in a straightforward and natural way.
The second stage of the work involved developing the actual \pybind-based binding
code to expose the \pressio~functionalities to Python.
This stage benefited from the ease of use and power of \pybind,
which is written to naturally target C++ and, in most situations,
leading to simple binding code.
This is what we observed for \pressio. For example, the binding
code to expose the basic LSPG functionalities is less than $100$ lines.

\lstset{style=pystyle}
\begin{lstlisting}[
    float,
    caption={Sample Python adapter class satisfying the API for \pressioFpy.},
    label={lst:pythonAdapter}]
class Adapter:
  def __init__(self, *args):
    # initialize (if needed)
    # create velocity vector, f, and Jacobian matrix, Jac

  # compute velocity, f(x,t;...), for a given state, x
  def velocity(self, x, t):
    # compute f (here f is a member of the class)
    return self.f

  # given current state x(t):
  # 1. compute the spatial Jacobian, df/dx
  # 2. compute A=df/dx*B, B is typically a skinny dense matrix
  def applyJacobian(self, x, B, t):
    # compute Jac = df/dx (here Jac is a member of the class)
    # Jac is typically sparse, so we use Jac.dot(B)
    # When Jac is dense, use np.matmul(Jac,B)
    return self.Jac.dot(B)


\end{lstlisting}
Listing \ref{lst:pythonAdapter} shows an example of a Python class
satisfying the API for \pressioFpy, revealing clear similarities with the C++ listing \ref{lst:adapter}.
When instantiating a ROM problem, the user can decide to
provide an object with custom algebra operations (e.g., to specify
how to perform matrix-vector products) for \pressioFpy~to use where needed,
or let \pressioFpy~rely on default implementation of these operations.
In the latter scenario, \pressioFpy~operates
by making direct calls to NumPy and SciPy functionalities to act
directly on \pybindarr~data. This is enabled by importing the NumPy or SciPy
modules \textit{directly} inside the C++ code as \code{pybind11::objects}.
For example, if we need to compute a on-node matrix-vector product, we obtain optimal
performance by calling the Blas function \code{dgemv}
to operate directly on the underlying matrix and vector data.
One scenario where this is not entirely possible is when a user wants to run
a ROM with hyper-reduction. In this case, the user is required to provide
custom functionalities, namely the computation of the time-discrete operators.
Nothing else changes as far as matrix and vector manipulations since these low-level
operations are still handled by \pressioFpy.

\section{Results}
\label{sec:results}
This section aims at demonstrating the use of \pressio~in two different scenarios:
in \S~\ref{ssec:results_sparc}, we discuss the application of the LSPG ROM
to a large-scale simulation of a hypersonic-flow problem,
while in \S~\ref{ssec:results_p4py} we discuss overhead and performance of \pressioFpy.
\subsection{LSPG ROM for hypersonics}
\label{ssec:results_sparc}

This section presents representative results obtained
using SPARC (Sandia Parallel Aerodynamics and Reentry Code).
SPARC is a Computational Fluid Dynamics (CFD) code implemented in C++ for
solving aerodynamics and aerothermodynamics problems. It solves the compressible
Navier--Stokes and Reynolds-averaged Navier--Stokes (RANS) equations on structured and unstructured
grids using a cell-centered finite-volume discretization scheme, and
the transient heat and associated equations for non-decomposing and decomposing
ablators on unstructured grids using a Galerkin finite-element method \cite{howard2017SPARC}.
SPARC relies on the distributed numerical linear algebra data structures
in Tpetra which, in turn, leverages the Kokkos programming model to achieve performance
portability across next-generation computing platforms. The reader is referred to \cite{howard2017SPARC}
for details on the governing equations and code design of SPARC.

The version of SPARC we used for our test did not natively expose the
API needed by \pressio~shown in \ref{lst:adapter}. However, its modular design made
the implementation of an adapter class quite straightforward.
In fact, the ``velocity'' and the evaluation of the Jacobian functionalities
were already present within SPARC---used by the native code within
each nonlinear solver iteration---despite being buried within a few abstraction layers.
Therefore, we just needed to create an adapter class bypassing
these layers and accessing these functionalities directly.
Having access to the spatial Jacobian, computing the \code{applyJacobian}
required by \pressio~(see \ref{lst:adapter}) was straightforward because it simply involved calling
a method of the native Jacobian's data structure to compute its action on the operand $\bf B$.
Overall, we can say that implementing the adapter class was quite easy,
requiring no modifications to the core SPARC code,
but just the addition of a few methods including those to initialize the discretization,
read in basis vectors, and post-process the ROM results.
This experience is an example of what was discussed in \S~\ref{ssec:pressio_api},
i.e.,~an application that, despite being complex, required only minor effort to interface with \pressio~due to its modular and extensible design.

\begin{figure}
  \centering
  \begin{minipage}{0.31\textwidth}
  \input{./anc/blottner_schematic}(a)
  \end{minipage}
  \begin{minipage}{0.31\textwidth}
  \includegraphics[width=0.8\textwidth, trim={10cm 1cm 0 0}, clip]{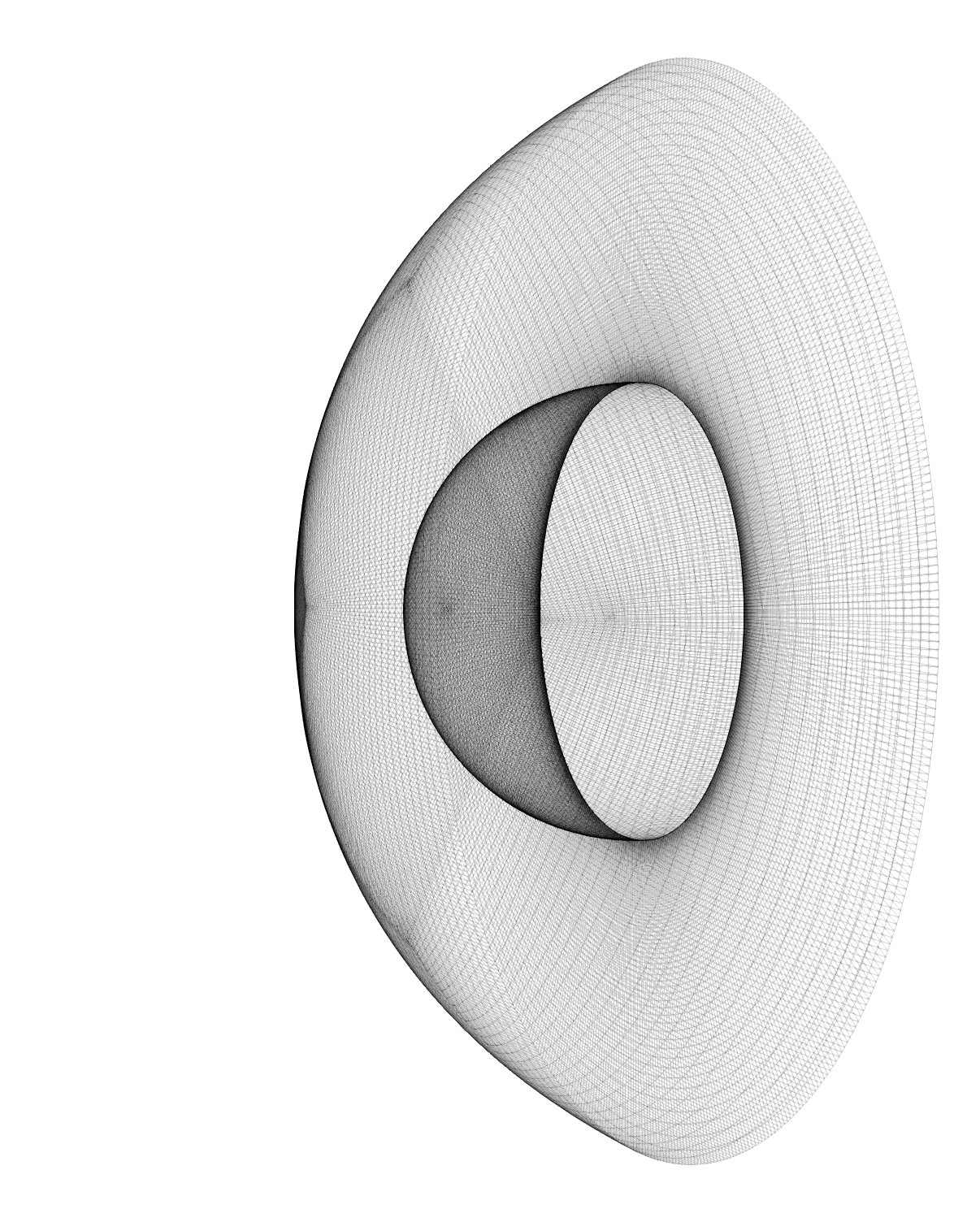}(b)
  \end{minipage}
  \begin{minipage}{0.31\textwidth}
  \includegraphics[width=0.8\textwidth, trim={10cm 1cm 0 0}, clip]{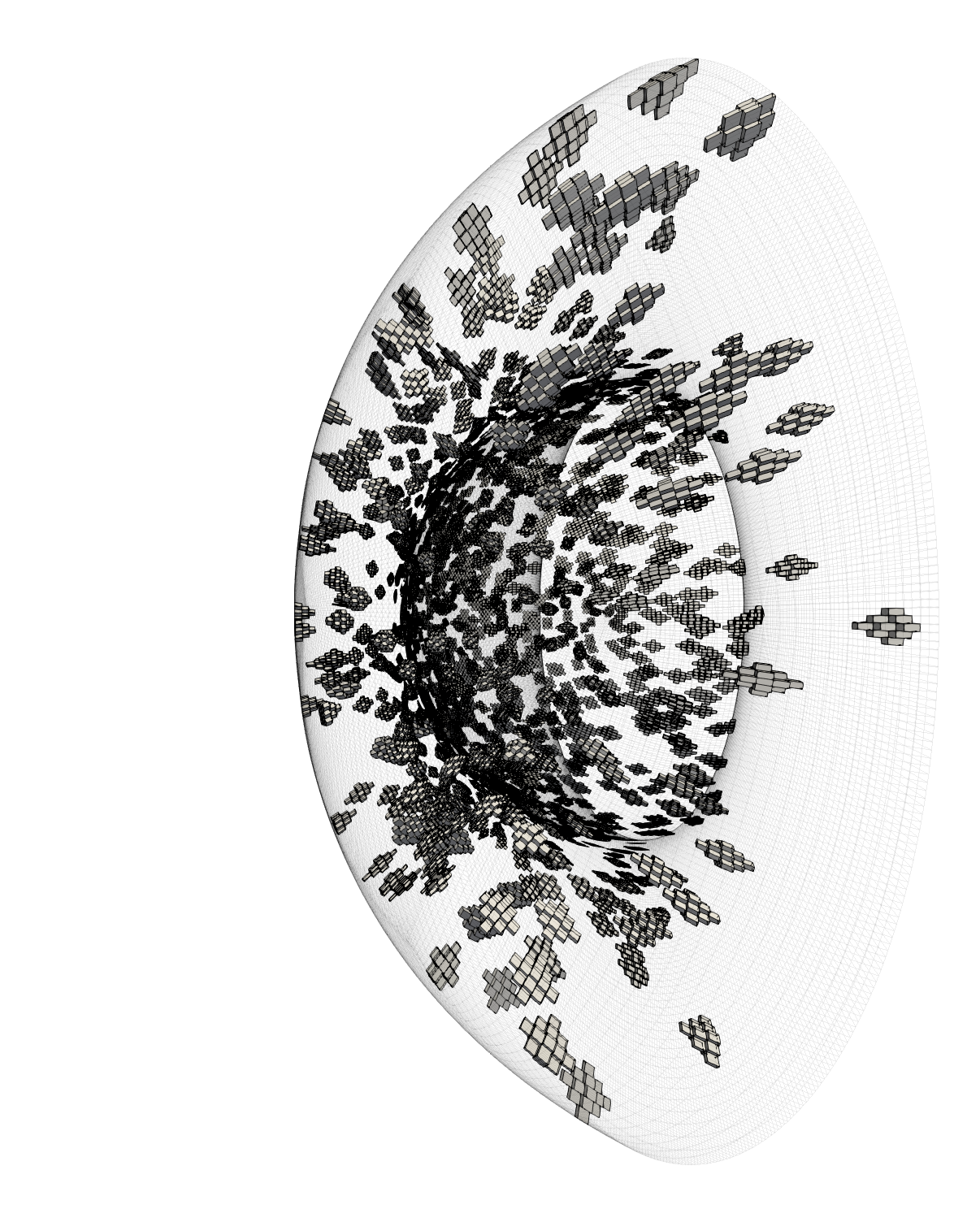}(c)
  \end{minipage}
  \caption{Panel (a): schematic of Blottner sphere,
    where $\rho_{\infty}$, $V_{\infty}$ and $T_{\infty}$ refer to the
    free-stream density, velocity and temperature, respectively. The computational domain is displayed with
    a dashed line. The bow shock to the left of the semi-sphere is where the hypersonic
    airflow almost suddenly decelerates to subsonic velocity, with density and temperature
    increasing near the sphere surface. Note that the schematic is shown in two dimensions
    for clarity, but the actual simulation is fully three-dimensional.
    Panel (b): angled front view of the full 3D mesh used for the full-order model of the Blottner sphere.
    Panel (c): angled front view of one of the sample meshes used to run the ROMs with
    hyper-reduction, which is composed
    of $1,678$ randomly selected cells in which the residual is sampled, along with neighboring
    cells and neighbors of neighbors, yielding $41,414$ cells in total,
    equivalent to $\sim 1 \%$ of the full mesh.}
 \label{fig:blottner_mesh}
\end{figure}

As a test case, we choose a CFD simulation of the Blottner sphere, a commonly used validation
case for hypersonic CFD solvers \cite{blottner1990sphere}.
It models a hypersonic flow impacting on a semi-sphere similar to the round tips commonly
found on hypersonic aircraft and spacecraft. A schematic of the problem is
shown in Figure~\ref{fig:blottner_mesh}~(a). The main feature is the bow shockwave just
upstream of the sphere, which can be difficult to capture accurately due to the shock
strength and a numerical phenomena know as the carbuncle problem \cite{maccormack2013carbuncle}.
Non-dimensional quantities characterizing the problem are the Mach number, $\Mach = 5.0$,
and the Reynolds Number, $\Rey=1,887,500$. Despite the large Reynolds number, this
problem is run without a turbulence model, and the boundary layer is
assumed to be laminar \cite{blottner1990sphere}.

The FOM is based on a spatial discretization using a structured mesh with $4,192,304$ cells
shown in Figure~\ref{fig:blottner_mesh}~(b). The mesh is designed so that the cells are well
aligned with the bow shock and the boundary layer on the sphere surface, and it extends upstream
enough to capture the bow shock but not so far so as to control the cell count.
Since the flow is modeled as laminar, there are $5$ conserved quantities, yielding the
total FOM degrees of freedom to be $\fomDof = 20,971,520$.
We focus on the pseudo-transient behavior of the system from a uniform $\Mach=5.0$ flow to steady state.
The pseudo time stepping is performed using the (implicit) backward Euler scheme with
fixed time step $\timeStep=2.5\times 10^{-6}$ seconds, yielding $64$ time instances.
At each time instance, the nonlinear system resulting from the implicit time
discretization is solved using a Newton--Raphson solver until the $\ell^2$-norm
of the residual is reduced by three orders of magnitude \cite{howard2017SPARC};
the initial guess is taken to be the solution at the previous time instance.
During the pseudo-transient phase, the bow shock
upstream of the sphere moves and deforms until it reaches a steady state.

We consider the LSPG method described in \S~\ref{ssec:romtheory_lspg}
with a linear trial subspace, and limit the study to a reproductive ROM.
This means that we fix the system parameters $\params$---which in this case
correspond to boundary conditions and physical constants---and assess the
ability of the ROM to reproduce the FOM results.
A predictive ROM study is outside the scope of the present work,
and we refer the reader to \cite{Blonigan2020} for a paper showing
ROMs developed in \pressio~yielding
good performance in a predictive setting for a hypersonic-flow problem.
To compute the basis matrix $\romBasis$, we take snapshots of the full-order
model state at all $64$ time instances, subtract
the initial condition $\initState(\params)$ from each snapshot, and compute the corresponding POD modes.
We consider four cases, namely $\dofRom = 8$, $16$, $32$, and $64$ modes.
For the ROM, we use the same time discretization method and time step size as the FOM.
The LSPG minimization problem in Eq.~\eqref{eq:disc_opt_problem} employs
$\stateRef = \initState$ and is solved using the Gauss--Newton method
solver implemented in \pressio~that assembles and solves the normal equations,
terminating when the $\ell^2$-norm of the normal-equations relative residual is reduced by three orders of magnitude; the initial guess is taken
to be the solution at the previous time instance.
We consider two different choices of the weighting matrix $\weightMat$ for the
problem in Eq.~\eqref{eq:disc_opt_problem}:
\paragraph{1. Diagonal matrix without hyper-reduction:\nopunct} we choose
$\weightMat=\weightMatDiag\in\mathbb{R}^{\fomDof\times\fomDof}$,
where $\weightMatDiag\in\mathbb{R}^{\fomDof\times\fomDof}$ is a diagonal matrix
whose elements correspond to the sizes of the control volumes in the
finite-volume discretization divided by the time step $\timeStep$.
We found that this choice of $\weightMat$ greatly improves the convergence
rate of Gauss--Newton iterations.
This case is useful for demonstration purposes but is expected not to yield
substantial computational savings because no hyper-reduction is employed.
For brevity, below we refer to this case with the abbreviation $\text{sm}_{100}$
to indicate that this uses $100\%$ of the full mesh.
\paragraph{2. Diagonal matrix with collocation hyper-reduction:\nopunct}
we choose $\weightMat=\collocMat\weightMatDiag\in\mathbb{R}^{z\times\fomDof}$,
where $\weightMatDiag\in\mathbb{R}^{\fomDof\times\fomDof}$ is the same as the case above,
and the sample matrix $\collocMat\in\{0,1\}^{z\times\fomDof}$ corresponds to
$z(\ll\fomDof)$ randomly selected rows of the identity matrix.
Note that, as suggested in \cite{CarlbergGnat}, this random sampling is
done such that we obtain cells at each boundary.
To enable this in SPARC we use a sample mesh. The main difficulty of using a sample mesh
stems from the need to perform computations on its odd topology comprised of disjoint sets of cells.
However, this proved relatively easy in SPARC because of its support
for unstructured meshes---specifically those in
the EXODUS II\footnote{https://gsjaardema.github.io/seacas/html/index.html}
file format \cite{howard2017SPARC}.
Data structures for unstructured meshes are, in fact, well-suited to
handle a sample mesh (as anticipated in \S~\ref{ssec:pressio_hyperreduction}),
because they hold, for every cell of the domain, the full connectivity list
and faces corresponding to side sets. Therefore, SPARC did not have to be modified to perform
computations on a sample mesh. This confirms our statement in \S~\ref{ssec:pressio_hyperreduction}:
implementing hyper-reduction for unstructured meshes is, in general,
relatively straightforward.
In this work, we explore sample meshes corresponding to $1$\%, $2$\%, $5$\%, and $10$\% of the full mesh,
and we refer to them with the symbols $\text{sm}_1$, $\text{sm}_2$, $\text{sm}_5$ and $\text{sm}_{10}$, respectively.
A visualization of the sample mesh containing $1$\% of the full mesh is shown in Figure~\ref{fig:blottner_mesh}~(c).

All of the results presented below for the Blottner sphere are run on a Sandia cluster
containing a total of $1,848$ nodes with Infiniband interconnect, each node
having two 8-core Intel(R) Xeon(R) CPU E5-2670 0 @ 2.60GHz and 65GB of memory.
To run a FOM simulation, we use $512$ MPI processes, distributed over $32$ nodes.
This configuration follows SPARC's requirements
and allows us to balance the memory footprint of the application.
The hyper-reduced ROMs run on a single node, since the sample mesh is small enough.
The mesh for the ROM without hyper-reduction is distributed using the same number of nodes used for the FOM.
We choose Kokkos for the ROM data structures and computations. This is advantageous
for two main reasons: first, it yields seamless interoperability with the Tpetra-based
data structures of SPARC; second, it allows us, in the future, to tackle
more complex SPARC problems by leveraging
on-node threading models or directly on the GPU without modifying the code.
In this work, Kokkos is built with a serial backend.
A detailed study comparing various threading models
for running the ROM is left to future work.
\begin{figure}[!h]
  \centering
  \includegraphics[width=0.335\textwidth, trim={0 0cm 6cm 0}, clip]{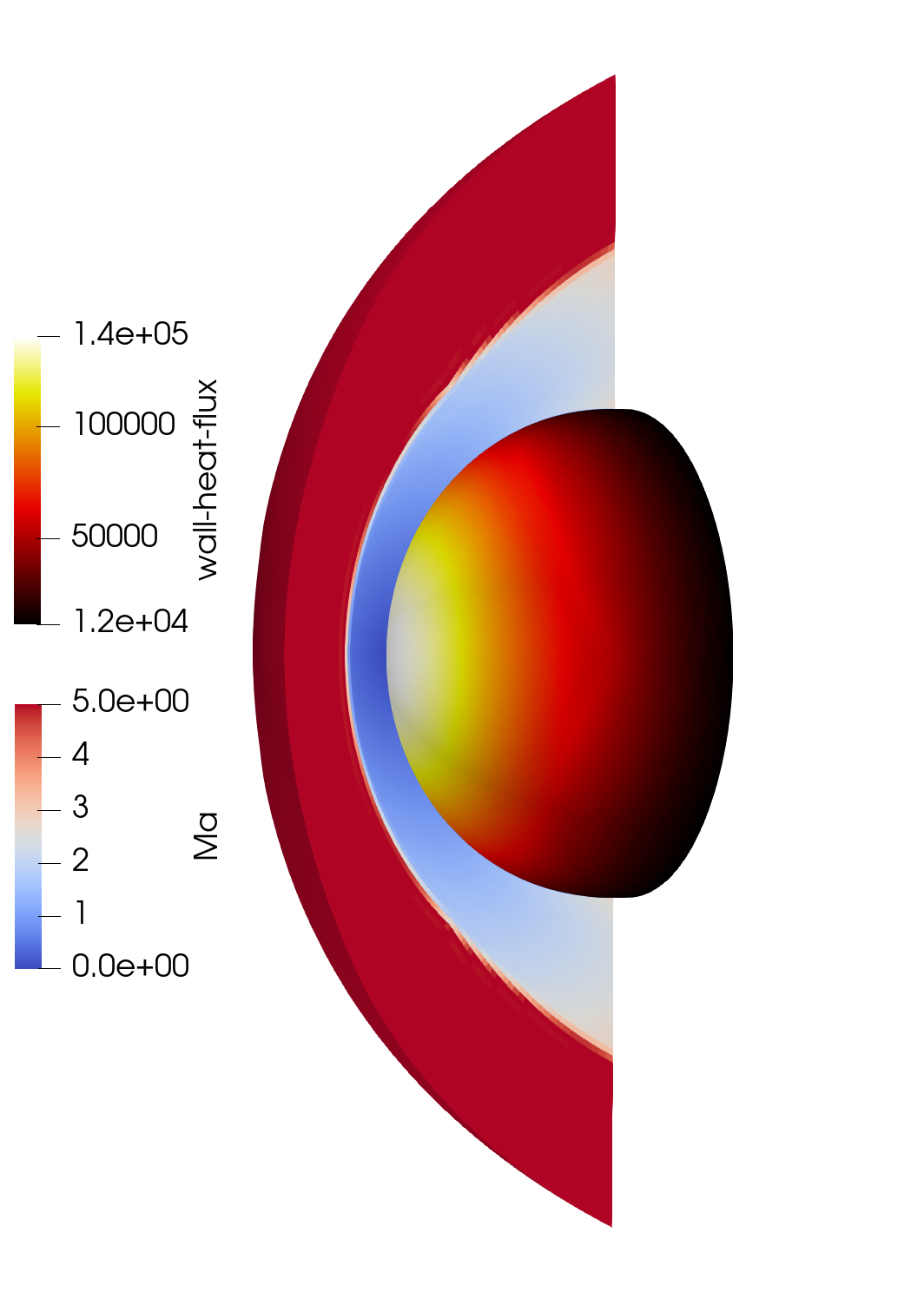}(a)
  \includegraphics[width=0.24\textwidth, trim={3cm 0cm 2cm 0}, clip]{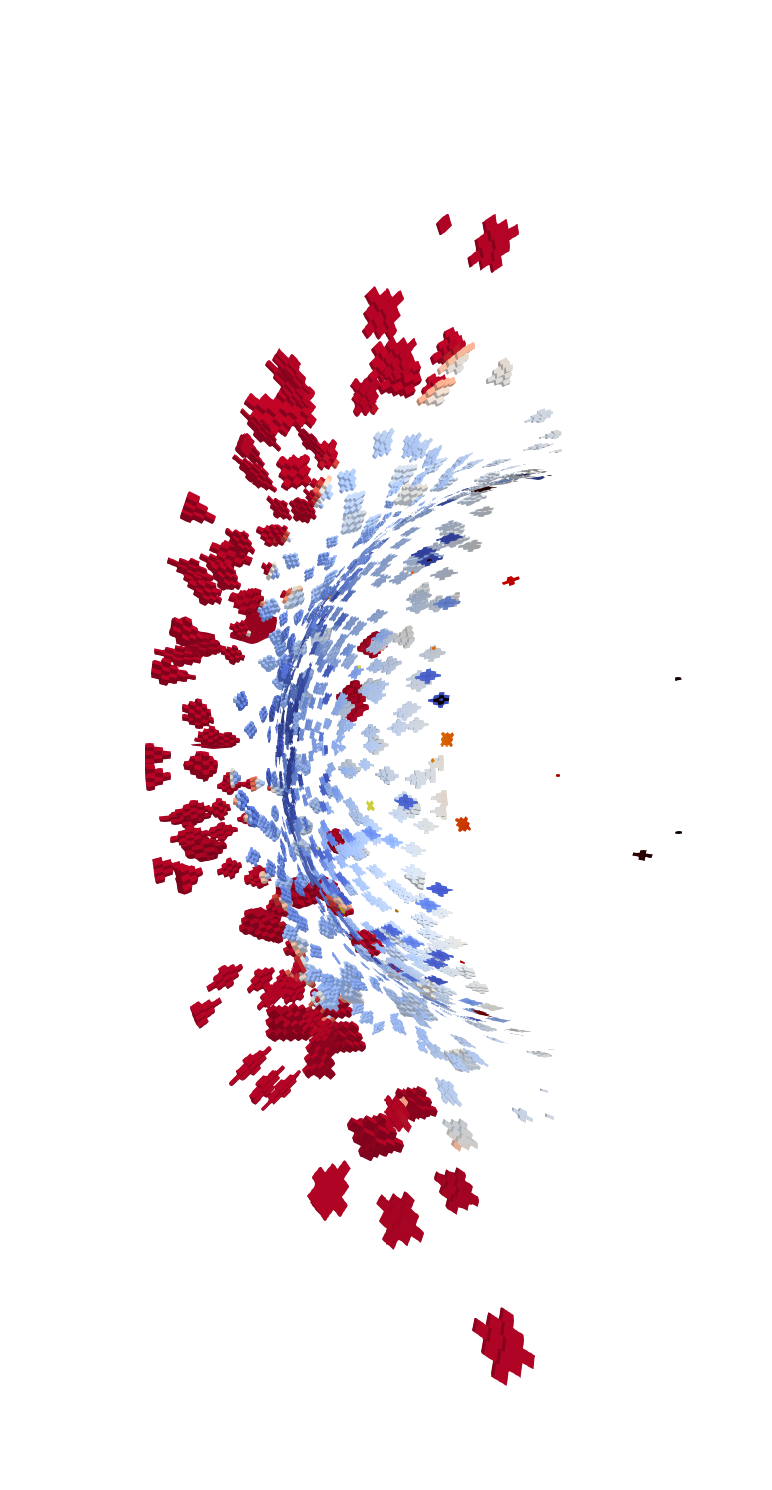}(b)
  \includegraphics[width=0.24\textwidth, trim={3cm 0cm 2cm 0}, clip]{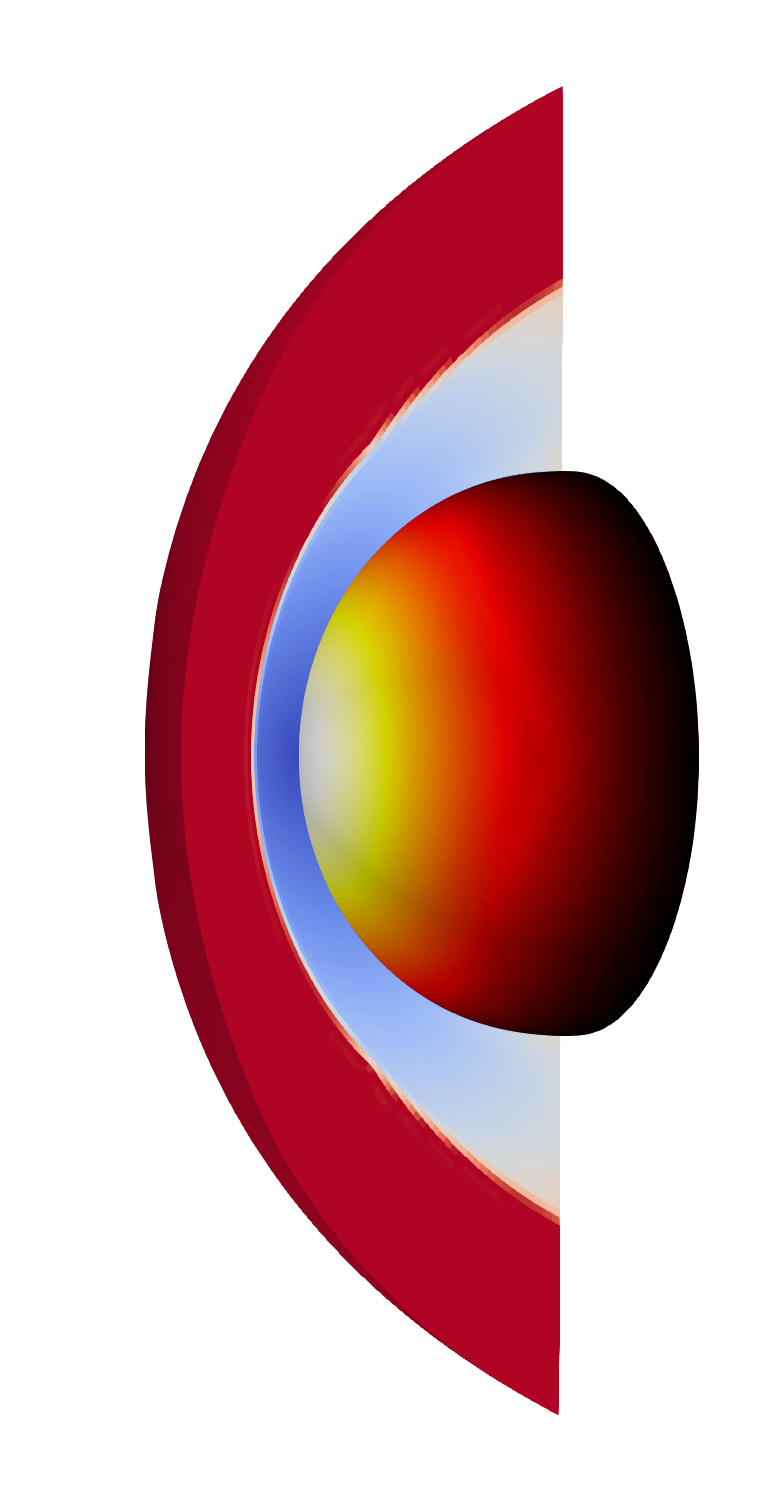}(c)
  \caption{Three-dimensional visualizations of the Blottner sphere problem
    at the final time step (i.e.,~steady state) displaying the flow color-coded
    by the Mach number ($\Mach$) and the sphere surface color-coded by the local heat flux.
    Panel~(a) shows the results from a full-order model simulation.
    Panels~(b,c) show those obtained from the LSPG ROM with 8 POD modes and $1$\% sample mesh:
    in (b) we show the sample mesh only, and in (c) we plot the
    ROM-based solution reconstructed over the full mesh.
  }
  \label{fig:mach_heat_3d}
\end{figure}

\paragraph{Accuracy:\nopunct}
Figure~\ref{fig:mach_heat_3d} shows representative visualizations of the flow,
color-coded by the Mach number, and the sphere surface,
color-coded by the local heat flux.
The plots show that the ROM faithfully captures the bow shock
and high heating around the flow stagnation point,
yielding differences with the FOM that are nearly indistinguishable.
To quantify these differences, we report in Table~\ref{tab:errors} the
relative errors of the ROM with respect to the FOM measured in
the discrete $\ell^2$- and $\ell^\infty$-norms for the state vector
on the entire computational domain, and vectors containing
the wall heat flux and wall shear stress at all surface cell faces.
For each sample mesh except $\text{sm}_1$, the error decreases
monotonically with the reduced dimension $\dofRom$ (i.e., the number of basis
vectors). We emphasize that as the reduced dimension increases,
monotonic decrease in the errors reported here is not guaranteed,
even when full mesh sampling is used.
Overall, we can say that for all configurations explored, the ROM
accurately reproduces the FOM results.
Since this is a reproductive ROM, when we use the full set
  of POD modes, $p=64$, one would expect to recover the FOM
  state committing an error that is close to machine zero.
  For this test case, this does not occur because, as anticipated before,
  the convergence criterion for both the FOM and ROM nonlinear solvers is not
  the absolute norm of the residual reaching machine zero, rather a reduction
  of its relative norm. Hence, this tolerance poses a limitation.
  We remark that a convergence criterion based on the relative residual
  norm reduction is typical for large-scale CFD problems,
  since obtaining machine zero is generally unfeasible.

\begin{table}
  \centering
\begin{tabular}{l|l|cc|cc|cc}
  \multirow{2}{*}{} & \multirow{2}{*}{$p$} & \multicolumn{2}{c|}{state error} &
	\multicolumn{2}{c|}{heat-flux error} & \multicolumn{2}{c}{wall-shear error} \\
  & & $\ell^2$-norm & $\ell^{\infty}$-norm & $\ell^2$-norm &
	$\ell^{\infty}$-norm  & $\ell^2$-norm & $\ell^{\infty}$-norm \\
  \midrule
  \multirow{4}{*}{\rotatebox{90}{$\text{sm}_1$}}
  &  08 & 5.013e-04 & 5.621e-03 & 2.912e-03 & 3.581e-03 & 1.644e-03 & 3.586e-03 \\
  &  16 & 8.354e-05 & 9.039e-04 & 1.838e-04 & 4.528e-04 & 1.796e-04 & 6.041e-04 \\
  &  32 & 2.807e-05 & 1.850e-04 & 2.193e-04 & 3.115e-04 & 1.517e-04 & 3.369e-04 \\
  &  64 & 1.617e-04 & 2.412e-03 & 7.174e-04 & 1.086e-03 & 4.218e-04 & 8.570e-04 \\
  \midrule
  \multirow{4}{*}{\rotatebox{90}{$\text{sm}_2$}}
  &  08 & 4.396e-04 & 5.478e-03 & 1.594e-03 & 3.242e-03 & 1.055e-03 & 3.478e-03 \\
  &  16 & 9.335e-05 & 9.974e-04 & 7.033e-04 & 1.105e-03 & 4.197e-04 & 1.040e-03 \\
  &  32 & 1.114e-05 & 1.098e-04 & 9.334e-05 & 1.529e-04 & 4.903e-05 & 1.273e-04 \\
  &  64 & 4.603e-06 & 5.229e-05 & 2.660e-05 & 4.312e-05 & 1.555e-05 & 4.300e-05 \\
  \midrule
  \multirow{4}{*}{\rotatebox{90}{$\text{sm}_5$}}
  &  08 & 4.404e-04 & 5.210e-03 & 1.754e-03 & 2.824e-03 & 9.739e-04 & 2.741e-03 \\
  &  16 & 1.356e-04 & 1.016e-03 & 1.196e-03 & 1.844e-03 & 7.411e-04 & 1.786e-03 \\
  &  32 & 1.000e-05 & 1.076e-04 & 9.340e-05 & 1.461e-04 & 4.829e-05 & 1.177e-04 \\
  &  64 & 1.564e-06 & 2.426e-05 & 8.238e-06 & 1.429e-05 & 4.649e-06 & 1.325e-05 \\
  \midrule
  \multirow{4}{*}{\rotatebox{90}{$\text{sm}_{10}$}}
  &  08 & 5.318e-04 & 4.589e-03 & 2.066e-03 & 3.460e-03 & 1.184e-03 & 3.178e-03 \\
  &  16 & 9.858e-05 & 9.816e-04 & 7.987e-04 & 1.228e-03 & 4.781e-04 & 1.159e-03 \\
  &  32 & 7.368e-06 & 9.632e-05 & 6.368e-05 & 1.025e-04 & 3.205e-05 & 7.851e-05 \\
  &  64 & 9.215e-07 & 1.520e-05 & 2.509e-06 & 4.087e-06 & 1.840e-06 & 4.707e-06 \\
  \midrule
  \multirow{4}{*}{\rotatebox{90}{$\text{sm}_{100}$}}
  &  08 & 5.073e-04 & 4.563e-03 & 2.048e-03 & 2.865e-03 & 1.162e-03 & 2.546e-03 \\
  &  16 & 8.075e-05 & 1.033e-03 & 6.144e-04 & 9.334e-04 & 3.705e-04 & 9.445e-04 \\
  &  32 & 3.753e-06 & 1.036e-04 & 1.257e-05 & 2.557e-05 & 6.534e-06 & 2.187e-05 \\
  &  64 & 9.849e-07 & 7.747e-06 & 4.059e-06 & 6.564e-06 & 3.121e-06 & 8.260e-06 \\
\end{tabular}
\caption{Relative norms of the errors for target observables between the ROM prediction and
  the full-order model results of the Blottner sphere problem. The metrics are computed at steady state.}
\label{tab:errors}
\end{table}

\begin{figure}[!t]
  \centering
  \includegraphics[width=0.7\linewidth]{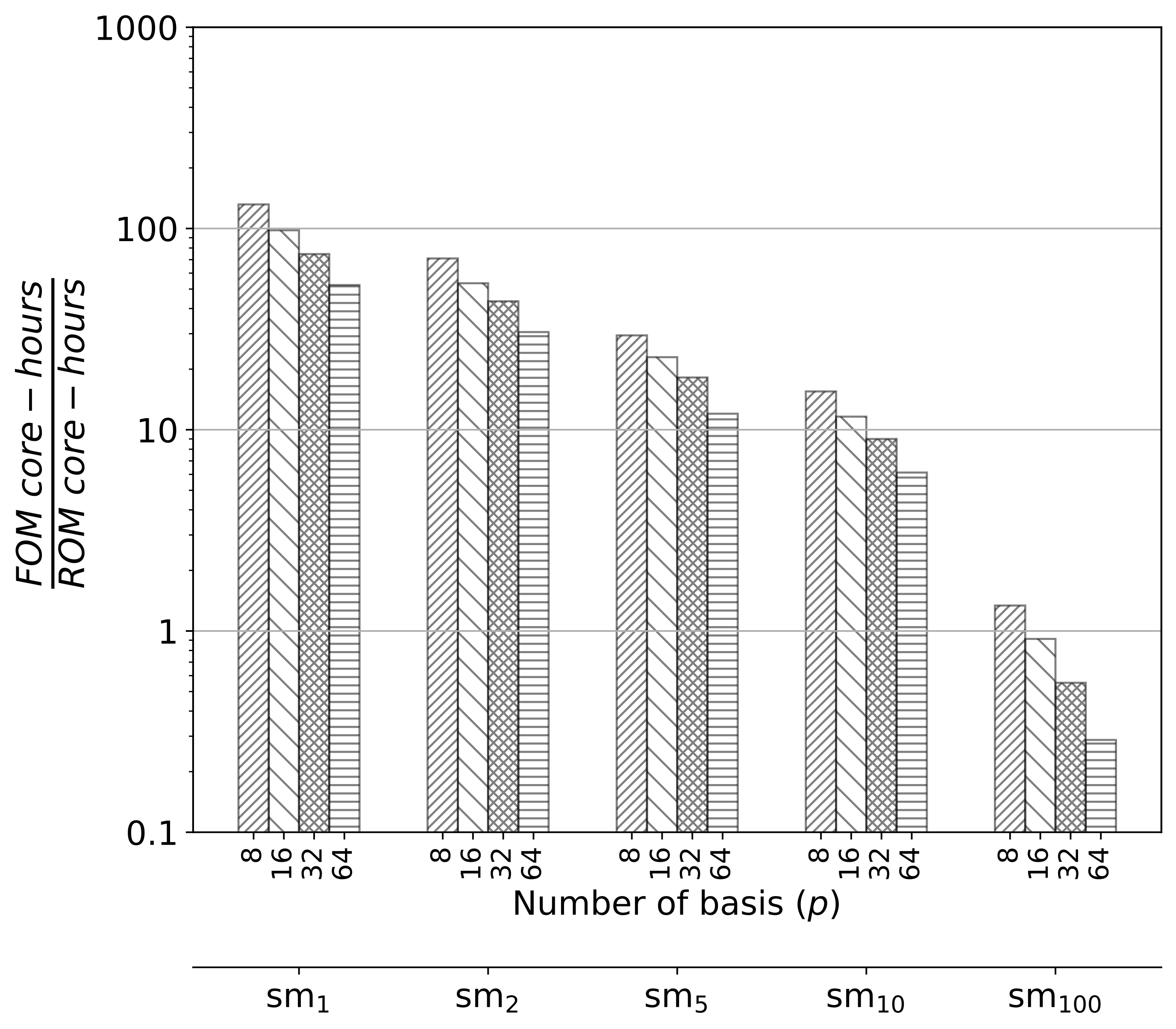}
  \caption{Performance results for the Blottner sphere test case showing the core-hours reduction
    factor of the LSPG ROM with respect to the full order model.
    Each data point is obtained by computing the geometric mean of a set of three runs.
    The hyper-reduced ROM cases (identified by labels $\text{sm}_1$, $\text{sm}_2$,
    $\text{sm}_5$ and $\text{sm}_{10}$) are run on a single node, while the
    ROM without hyper-reduction ($\text{sm}_{100}$) uses 32 nodes to distribute the mesh.}
  \label{fig:sparc_timing}
\end{figure}

\paragraph{Performance:\nopunct}
Figure~\ref{fig:sparc_timing} summarizes performance results in terms
of the computational cost reduction factor of the LSPG ROM with respect to the FOM,
computed as the ratio between the FOM core-hours and ROM core-hours.
The results are grouped by the sample mesh size.
The computational savings for the hyper-reduced ROMs are substantial. The largest speed up ($>100X$) is
obtained with the $1\%$ sample mesh and $\dofRom=8$, while the smallest ($\sim 5X$)
is obtained with the $10\%$ sample mesh and $\dofRom=64$.
It is also interesting to note that the ROM {\it without} hyper-reduction
yields a gain when using $\dofRom=8$. As expected, if the reduced system is
large enough and no hyper-reduction is used, the cost of the ROM is larger
than the FOM, as seen for sm$_{100}$ and $\dofRom\in{16,32,64}$.

\subsection{\pressioFpy~vs.~pressio native C++}
\label{ssec:results_p4py}

In this section, we discuss the performance and overhead incurred
when using \pressioFpy~rather than \pressio~directly from C++.
We consider a reproductive scenario with an inviscid Burgers problem written
in the conservative form
\begin{align}
  \frac{\partial u(x,t)}{\partial t} +
  \frac{1}{2} \frac{\partial (u^2(x,t))}{\partial x}
  & = \alpha \exp(\beta x), \notag \\
  u(0,t) &= \gamma, \forall t > 0, \label{eq:burgers}\\
  u(x,0) &= 1, \forall x \in [0, 100], \notag
\end{align}
where $u$ is the state variable, $x$ represent the spatial variable, $t$ is time,
and $\alpha$, $\beta$, $\gamma$ are real-valued parameters chosen as
$\alpha=\beta=0.02$ and $\gamma=5$.
We use a finite-volume spatial discretization with $\fomDof$ cell
of uniform width and Godunov upwinding.
The semi-discrete version of the problem above takes the form of Eq.~\eqref{eq:govern_eq}.
We explore full-order models of varying dimension $\fomDof = 2^{k}$,
with $k\in\{10, 11, 12, 13, 14, 15\}$.
Below, we purposefully limit our attention to performance and overhead, and
omit a discussion of the accuracy of the ROM which, for the problem above,
can be found in, e.g., \cite{deeprom}.

We have developed two implementations of the full-order problem in \eqref{eq:burgers},
one in Python that meets the API shown in listing~\ref{lst:pythonAdapter}, and one in C++ using
Eigen\footnote{http://eigen.tuxfamily.org} data structures that meets
the API shown in listing~\ref{lst:adapter}.
The Python implementation uses NumPy and SciPy, and is optimized with Numba decorators to translate
target loops---in the velocity and Jacobian calculation---to optimized machine code at runtime.
We use OpenBLAS~$0.3.7$\footnote{https://www.openblas.net/} built with GCC~$9.2.0$
as a linear algebra backend for both NumPy/SciPy and Eigen ensuring that
all computations involving dense operators call the same BLAS/LAPACK routines at the lowest level.

We focus on two reduction methods: Galerkin projection as
in \eqref{eq:cont_rom} using a 4th-order explicit Runge--Kutta time integrator
and LSPG \eqref{eq:disc_opt_problem} with a backward Euler time stepper.
For both Galerkin and LSPG, we consider ROMs of dimension $\dofRom \in\{ 32, 64, 128, 256\}$.
In all cases, we use $\weightMat = \identity$ (the benefits of hyper-reduction
are outside the scope of this section), use the initial condition $u(x,0) = 1$
as reference state, and employ a linear trial subspace for projection.
To compute the basis for Galerkin, we run the full-order model over
the interval $t \in [0, 2.048]$ using a 4th-order Runge--Kutta scheme,
collecting the snapshot matrix, subtracting the initial condition $u(x,0)=1$
from each snapshot and computing the POD.
To compute the basis vectors for LSPG, we use the same procedure except that
a backward Euler time scheme is used over a time interval $t \in [0, 0.512]$.
The time interval chosen for LSPG is shorter than the one for Galerkin
because LSPG is more expensive, and this setup allows us to have a reasonable
turnaround time for both.
In all cases, the resulting total number of steps can be exactly factorized
by any of the selected reduced dimensions $\dofRom$, and we use a
time step $\Delta t = 5\times 10^{-4}$, ensuring stability for all simulations.

For LSPG, we solve the nonlinear least-squares problem at each time step
using a Gauss--Newton solver implemented in \pressio~that assembles and
solves the normal equations.
To solve the normal equations, we use the \code{getrf} direct LU-based solver from LAPACK.
In Python, one can directly call \code{getrf} via the low-level LAPACK
functionalities in \code{scipy.linalg.lapack}.
In Eigen, built with LAPACK support, \code{getrf} is the implementation
underlying the LU decomposition of a dense matrix.

For LSPG, an important detail is the data structure type used for
the spatial Jacobian arising from the semi-discrete formulation of the system in \eqref{eq:burgers}.
In general, spatial Jacobian operators for problems involving local information
(e.g., finite differences and finite volumes) are sparse by definition
and, thus, typically represented as sparse matrices due to the smaller memory footprint.
For the purposes of this analysis, one caveat is that using a sparse Jacobian
implies using the native implementations in SciPy and Eigen, since we cannot rely on a common backend.
For a dense Jacobian, instead, as anticipated before, we can rely on the same BLAS backend.
To this end, we developed two versions of the spatial Jacobian for problem \eqref{eq:burgers},
one relying on a sparse and one on a dense matrix representation.
This allows us to quantify the overhead of \pressioFpy~with and without
biases due to different implementations. For all dense operators, we use
column-major ordering, yielding seamless compatibility
with the native Fortran API of BLAS/LAPACK, while for the sparse Jacobian we use the
Compressed Sparse Row (CSR) format.

\begin{figure}[!t]
  \centering
  \includegraphics[width=0.43\linewidth]{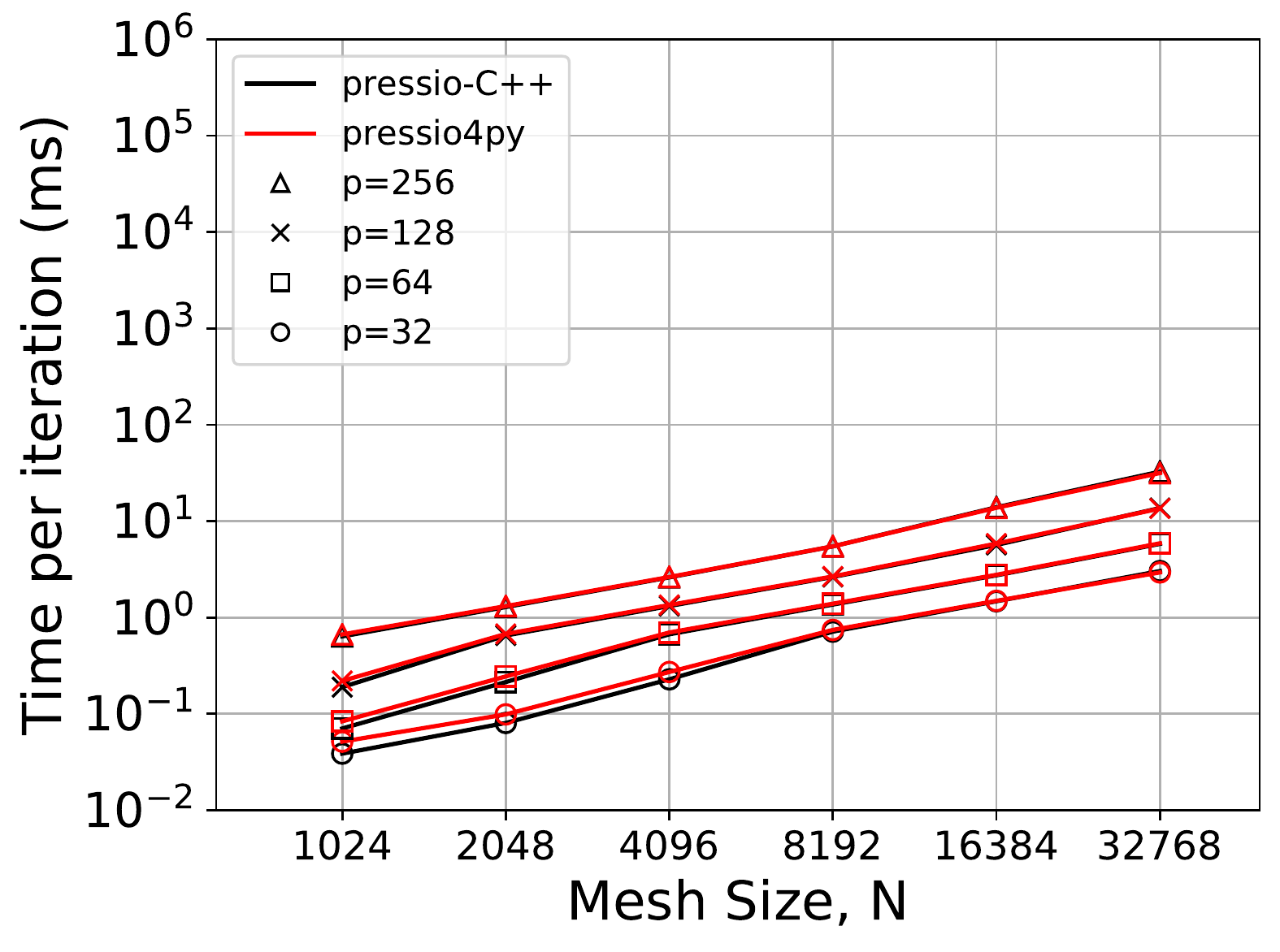}(a)\\
  \includegraphics[width=0.425\linewidth]{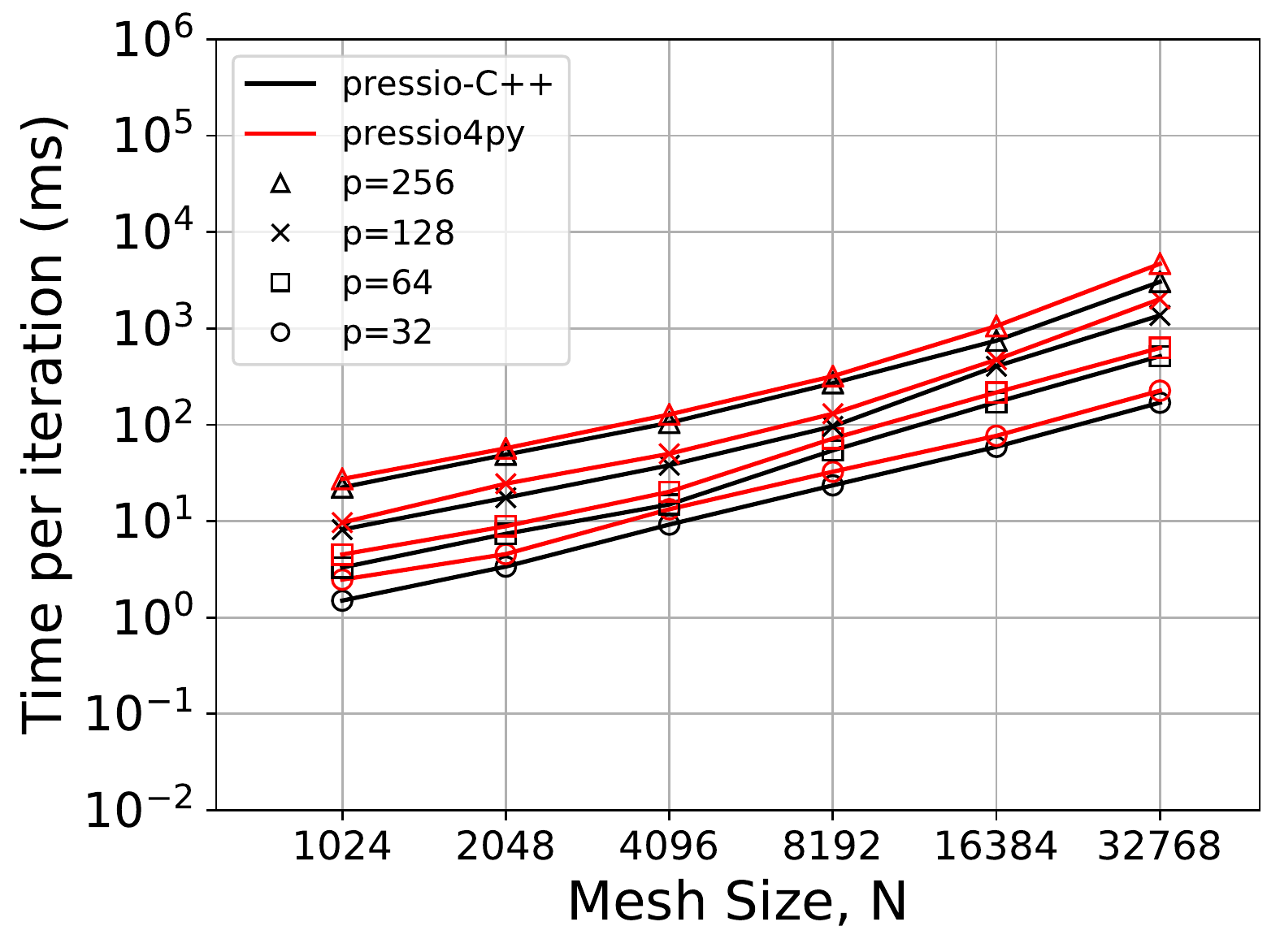}(b)
  \includegraphics[width=0.425\linewidth]{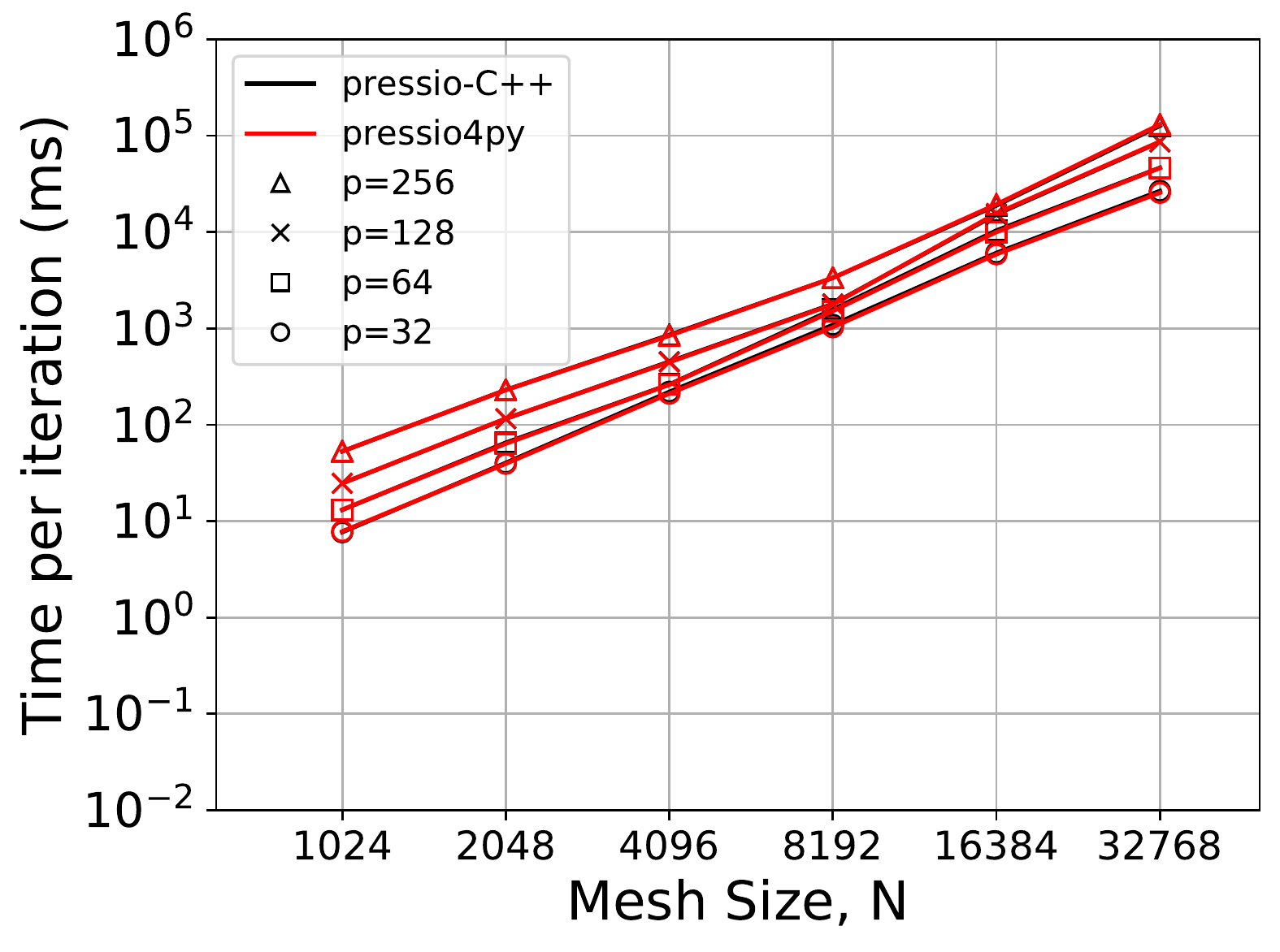}(c)
  \caption{Figure~(a): time per iteration (milliseconds) as a function of the mesh size
    obtained for Galerkin ROM applied to problem \eqref{eq:burgers} using the native
    C++ in \pressio~and \pressioFpy~for four different ROM sizes.
    Figure~(b) shows the corresponding results obtained for LSPG using a {\it sparse}
    matrix for the spatial Jacobian in the Burgers implementation, while (c) shows those obtained
  using a {\it dense} Jacobian matrix.}
  \label{fig:p4py_burgers}
\end{figure}

Figure~\ref{fig:p4py_burgers} shows the results obtained for Galerkin~(a),
and LSPG when using a sparse~(b) and dense~(c) matrix for the application's spatial Jacobian.
Each plot displays, in a semi-log scale, the time per iteration (in milliseconds) as a function of the mesh size,
obtained with the native C++ in \pressio~and \pressioFpy~for selected values of the ROM size, $\dofRom$.
Each data point is the geometric mean of a sample of $10$ replica runs, and error bars
are omitted because the statistical variability is negligible.
All runs were completed as single-threaded processes on a dual socket machine, containing two
18-core Intel(R) Xeon(R) Gold 6154 CPU @ 3.00GHz and 125GB of memory.

The plots show that Galerkin has the smallest time cost per iteration,
while LSPG based on a dense Jacobian has the largest time per iteration.
Figure~\ref{fig:p4py_burgers}~(a) shows that for Galerkin, up to about $\fomDof = 4096$,
the C++ version perform slightly better than the Python one. However, as the mesh size
increases, the two codes exhibit nearly identical performance.
As expected, panel~(b) shows that for LSPG with a sparse spatial Jacobian, the differences
between C++ and Python are more evident, revealing the bias due to the different implementation.
Panel~(c) shows that when LSPG is run with a dense spatial Jacobian, the differences
between the two codes are nearly indistinguishable. This analysis demonstrates that
the overhead incurred due to the Python layer of \pressioFpy~is effectively negligible.

\section{Conclusions}
\label{sec:conclusions}

This article presented \pressio, an open-source project aimed at enabling leading-edge
projection-based reduced order models (ROMs) for large-scale nonlinear dynamical systems.
We discussed the algorithms currently supported and those planned for upcoming releases,
including the support for both linear and nonlinear trial manifolds, and the applicability
to reduce both spatial and temporal degrees of freedom for any system expressible
as a parameterized set of ODEs.

We described the minimal API required by \pressio, which is a natural choice for dynamical systems.
To give an glimpse of the software design choices behind \pressio, we highlighted
the use and benefits of generic programming.
We discussed how the core C++ implementation of \pressio~is complemented
with Python bindings to expose these functionalities to Python users
with negligible overhead and no user-required binding code.

We presented our experience with a production-level aerodynamic code,
highlighting the minimal additions to the native code needed
to construct the sample mesh and apply the LSPG ROM.
We showed the substantial computational performance gains obtainable using a sample mesh,
and discussed how this should be evaluated in the context of a tradeoff between performance and accuracy.
We then showed an example using the Python bindings library, \pressioFpy,
which has a negligible overhead compared to using the core C++ \pressio.

We designed and implemented \pressio~to be usable for both large-scale simulations, where performance
is critical, and prototyping problems, where the focus is on assessing the accuracy of a new ROM technique.
We are working on expanding the set of demonstration cases to explore advantages and
disadvantages of ROMs for various types of problems, further improvements to performance,
investigating various sample mesh techniques, and providing a default, automated functionality to
run nonlinear ROMs based on convolutional autoencoders.



\section*{Acknowledgments}
The authors thank Matthew Zahr for extremely helpful and insightful
conversations that led to the inception of this project. The authors also
thank Matthew David Smith, and Mark Hoemmen for helpful suggestions and feedback. This paper
describes objective technical results and analysis. Any subjective views or
opinions that might be expressed in the paper do not necessarily represent the
views of the U.S. Department of Energy or the United States Government.
Supported by the Laboratory Directed Research and Development program at
Sandia National Laboratories, a multimission laboratory managed and operated
by National Technology and Engineering Solutions of Sandia, LLC., a wholly
owned subsidiary of Honeywell International, Inc., for the U.S.\ Department of
Energy's National Nuclear Security Administration under contract DE-NA-0003525.
SAND2020-1279 J and SAND2020-1445 J.

\bibliographystyle{siamplain}
\bibliography{references}
\end{document}


\maketitle

\section{Software design overview}
\label{ssec:pressio_softwaredesign}

In this section, we provide an overview of key software engineering
aspects of \pressio~aiming
at (a) describing how we abstract mathematical concepts into classes;
(b) providing a coarse-grained overview of the various components and their connections;
and (c) explaining the design choices behind the API required by \pressio.
To convey the main ideas clearly and concisely, we describe at a high level
representative classes needed for a Galerkin ROM
with an explicit time integration, deliberately omitting overly fine-grained
details about classes, methods and syntax, which we believe would be
more distracting than useful. We refer the reader interested in
the full implementation details to the source code of \pressio~and its documentation directly.

\begin{figure}
  \centering
  \resizebox{0.9\columnwidth}{!}{
    \input{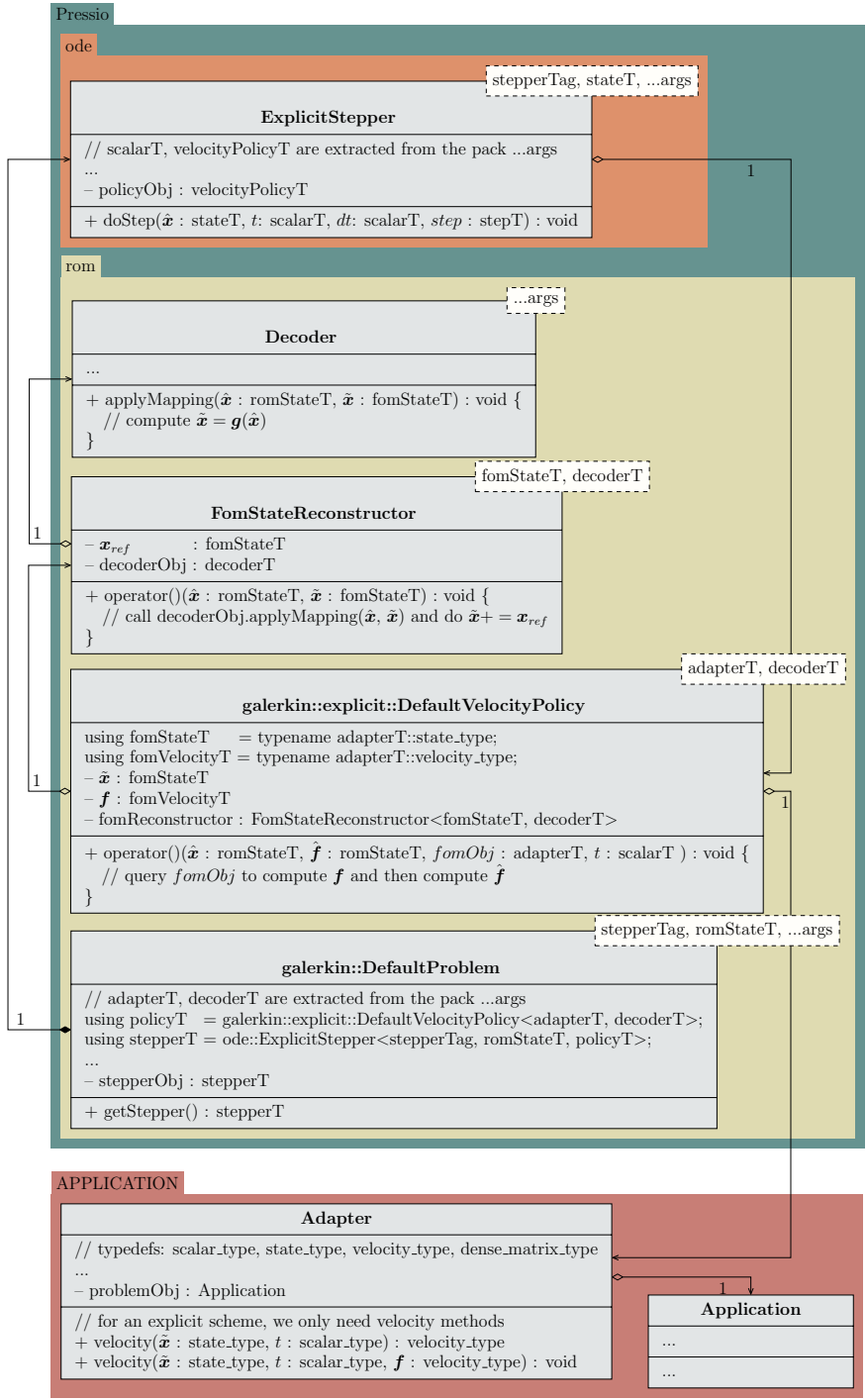}
  }
  \caption{UML class diagram of the main classes defining a Galerkin ROM
  using an explicit time stepping scheme.
  For the sake of readability, we adopt a few simplifications: (a) wherever possible
  we use the same notation used in \S~2 in the paper; (b) we omit base classes;
  (c) as non-standard extensions to the UML language, we use angle brackets $<>$ to represent templates,
  and the C++ keyword \code{using} to represent type aliases in classes, respectively.
  These minimal simplifications allow us to make the diagram substantially
  more readable and to highlight the interdependencies in a clearer way.}
\label{galerkinuml}
\end{figure}

Figure~\ref{galerkinuml} shows a (simplified) UML (Unified Modeling Language) class
diagram of the main classes needed for a Galerkin ROM using an explicit time stepping scheme.
For the sake of readability, wherever possible, we denote arguments and variables
with the same notation used in \S~2 of the paper.
A similar schematic for LSPG can be found in~\ref{sec:appendixB}.

To discuss the UML diagram in \ref{galerkinuml}, we start from the \code{DefaultProblem} class.
This class template represents the default Galerkin problem,
namely~Eq.~2.6 in the paper with $\weightMat = \identity$.
It needs as template arguments a (tag) type \code{stepperTag},
and types for the ROM state, adapter class and decoder.
\code{stepperTag} is used for tag dispatching,
and can be chosen from a set of options,
e.g., \code{ode::explicitmethods::Euler} or \code{ode::explicitmethods::BDF2}.
The variadic nature of the class facilitates the user since the
template arguments can be passed in arbitrary order.
We use metaprogramming to expand the pack and detect the needed types
according to specific conditions. For example, a type is detected as
a valid adapter class type only if it satisfies the API discussed in listing~1 in the paper.
Other conditions are used to detect the state and decoder types.
If \pressio~does not find admissible template arguments, a compile-time error is thrown.
An object instantiated from the class template \code{DefaultProblem} is responsible
for constructing a policy of type \code{DefaultVelocityPolicy},
a stepper object of type \code{ExplicitStepper}, and a reconstructor object of type \code{FomStateReconstructor}.
The policy defines \textit{how} the right-hand-side (velocity) of the Galerkin
system is computed, the stepper defines
how to advance the system over a single time step, and the reconstructor holds the
information and data (i.e., the decoder) to map a given ROM state vector to a FOM model state.
Once the Galerkin problem is instantiated, the contained stepper object can
be used with one of the integrators available inside \pressio/\code{ode}.

The UML diagram highlights that the application \textit{only} interacts with the policy class.
The policy object, in fact, calls the \code{velocity} method inside the adapter,
which in turn holds all the information about the FOM problem.
Minimizing the number of direct connections between the application
and \pressio~yields a much leaner structure less prone to errors.
The policy-based design seems a suitable choice allowing us to
create new problems by only operating at a low level while keeping
the interaction between the various components and the workflow intact.
This is in fact how we support variations of the basic Galerkin:
different policy (and problem) classes can be implemented encapsulating
different ways to compute the right-hand-side
shown in the formulations in Eq.~2.6.

\subsection{Example main file for Galerkin ROM}
\label{sec:appendixA}

The UML diagram description above enabled us to reason
about the global structure and interdependencies between the classes.
This section presents a sample main file and a description
of how to create and run a Galerkin ROM. Listing \ref{lst:cppmaingalerkin}
shows the skeleton of a C++ main file to instantiate and run
a default Galerkin ROM with forward Euler. We highlight the following parts.
On line 10, the user creates
an object of the adapter class (which, underneath, is expected to create
a full problem of the target application).
A decoder object is created on lines 15--17.
In this case, the listing shows a linear decoder,
which is defined by its Jacobian.
On line 20, a reference state for the full order model is defined,
and on line 23 the ROM state is defined.
The actual Galerkin problem is instantiated on lines 26--30.
The reader can cross-reference this class and its template arguments
with the UML diagram in Figure~\ref{galerkinuml}.
Finally, after the problem is created, the actual time integration is
run on line 34. For simplicity, we show here how to integrate for a target number of steps,
but other integration options are available.
\pressio, in fact, is implemented such that the concept of a stepper (whose responsibility is to only
perform a single step) is separated from how the stepper is used to advance in time.
This allows us to easily support different integrators (e.g. supporting variable time steps,
running until a fixed time is reached, or until a user-provided condition is met).

\lstset{style=cppstyle}
\begin{lstlisting}[
float,
    caption={Sample C++ main to create and run (explicit) Galerkin using \pressio.},
    label={lst:cppmaingalerkin}]
int main(int argc, char *argv[]){
  using adapterT        = /*adapter class type*/;
  using scalarT         = typename adapterT::scalar_t;
  using fomStateT       = typename adapterT::state_t;
  using decoderJacT     = typename adapterT::dmatrix_t;
  using romStateT       = /*ROM state type*/;
  using namespace pressio;

  // create app or adapter object
  adapterT fomObj(argc, argv, /*other args needed*/);

  const std::size_t romSize = /*set the Rom size */;

  // create the decoder, e.g. linear decoder
  decoderJacT phi = /*load/compute the decoder's Jacobian*/;
  using decoderT  = rom::LinearDecoder<decoderJacT>;
  decoderT decoderObj(phi);

  // set the FOM reference state
  fomStateT xRef = /*load/compute the fom reference state*/;

  // define ROM state
  romStateT romState(romSize);

  // create the Galerkin problem
  using stepperTag = ode::explicitmethods::Euler;
  using rom::galerkin::DefaultProblem;
  using galerkinT = DefaultProblem<stepperTag, romStateT,
                                   adapterT, decoderT>;
  galerkinT galerkinProb(fomObj, xRef, decoderObj, romState);

  // advance in time (t0, dt, Nsteps must be defined somewhere)
  // or one can use a different integrator
  ode::integrateNSteps(galerkinProb, romState, t0, dt, Nsteps);

  // map the final ROM state to the corresponding fom
  auto xFomFinal = galerkinProblem.fomReconstructor()(romState);

  return 0;
}

\end{lstlisting}

\subsection{Software Design for LSPG}
\label{sec:appendixB}
\begin{figure}[!t]
  \centering
  \resizebox{\columnwidth}{!}{ \input{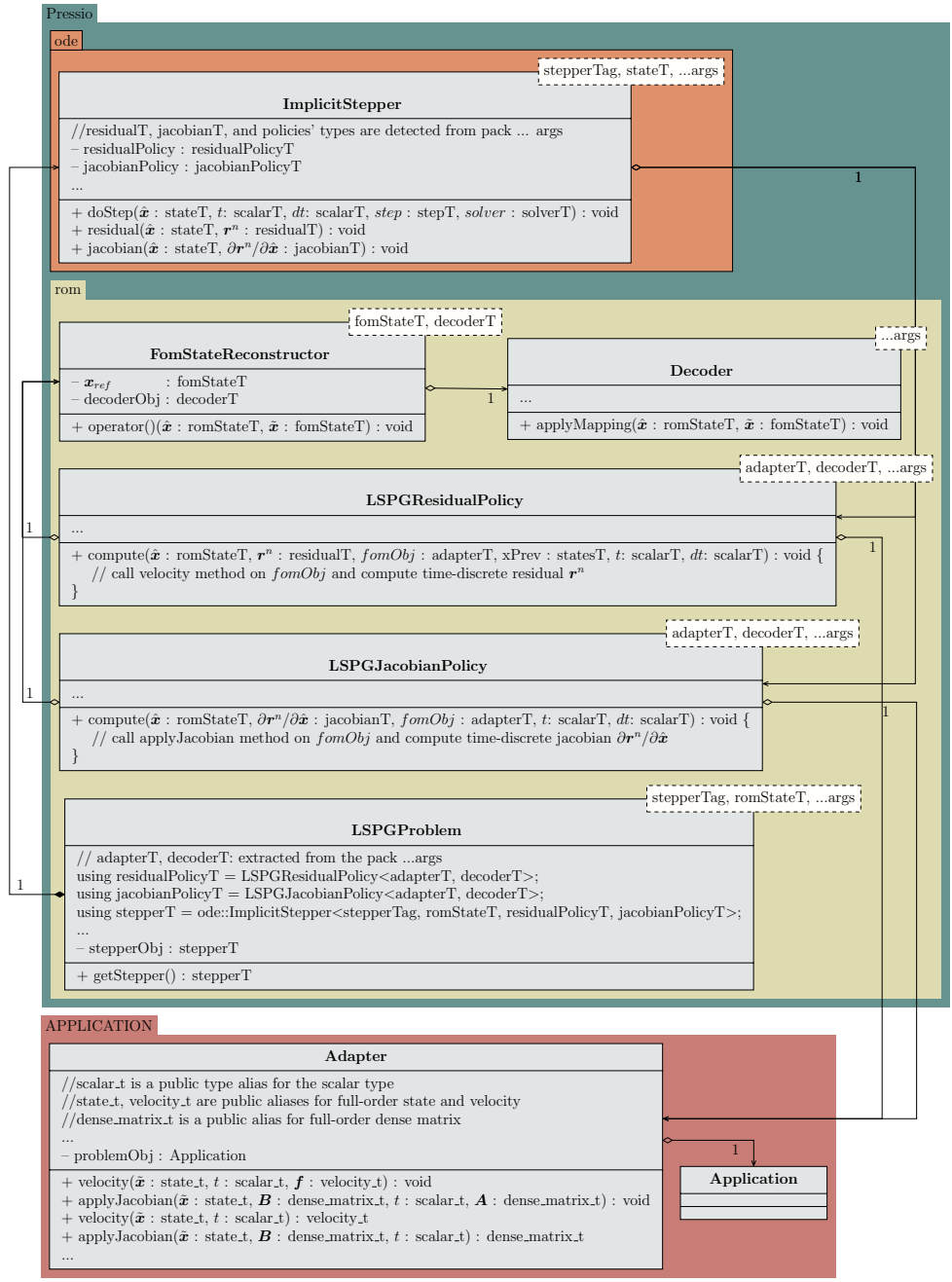} }
  \caption{UML class diagram of the basic structure of an LSPG problem.
  We use the same notation and simplifications as in Figure~\ref{galerkinuml}.}
\label{lspguml}
\end{figure}

Figure~\ref{lspguml} shows a simplified UML diagram of the main
classes composing an unsteady LSPG problem.
The \code{LSPGProblem} class is responsible for constructing
an object of the policy class \code{LSPGResidualPolicy},
one of the policy class \code{LSPGJacobianPolicy},
one of the stepper class \code{ImplicitStepper},
and one of the class \code{FomStateReconstructor}.
In this case, due to the implicit time integration,
we need two policies, one for the time-discrete residual
and one for the time discrete Jacobian.
The stepper defines how to advance the system over
a single time step, and the reconstructor holds the information and
data (i.e. the decoder) to map a given ROM state vector to a full-order model state.
Once the LSPG problem is created, the associated stepper object is
used to advance the ROM solution in time.
To this end, given the implicit nature of the stepper, a solver object
is needed as an argument to the \code{doStep} to solve the nonlinear system
at a given time step. For LSPG, the solver can be instantiated from one of the
least-squares solver classes inside the \pressio/\code{solvers}
currently supporting various types of Gauss-Newton nonlinear solvers,
e.g.~using the normal equations or QR factorization with an optional line search.
Alternatively, the user can provide a custom solver object.
At compile-time, \pressio~checks that the solver type is admissible by verifying
that it satisfies a specific API, e.g.~having a public
method \code{solve} accepting specific arguments.

The workflow is thus as follows: the time integrator (instantiated in the main, not shown here)
chosen by the user calls \code{stepperObj.doStep}, which triggers the solver via
\code{solver.solve}, which in turn triggers a callback
to \code{stepperObj.residual} and \code{stepperObj.jacobian}
to compute the time-discrete quantities.
The \code{residual} and \code{jacobian} methods in the \code{ImplicitStepper}
do not hold any implementation details, because they simply call
the \code{compute} method of the corresponding policy objects.
Similarly to the Galerkin problem discussed before, the classes for LSPG
are designed such that only the policies hold references to and interact with the application.
This policy-based design allows us to create various LSPG problems
by only changing the policies, without altering the overall workflow.

\clearpage
\section{Repositories and Versions}
The project and its components are accessible at:
\begin{itemize}
\item Project url: \hspace{2.2cm}\url{https://github.com/Pressio}
\item C++ source code: \hspace{1.2cm}\url{https://github.com/Pressio/pressio}
\item \pressioFpy~source code: \hspace{0.155cm}\url{https://github.com/Pressio/pressio4py}
\end{itemize}

\subsection{Code version used for \S~4.1}
The version of \pressio~used to generate the
results in \S~4.1 for the large-scale hypersonic
flow can be easily obtained since we used a Git \code{tag}
to mark the commit in our development.
The procedure is as follows:
\begin{enumerate}
\item Clone the core C++ \pressio~repository:\\
\code{\hspace{1cm}git clone https://github.com/Pressio/pressio.git}
\smallskip
\item Checkout the version used by doing:\\
\hspace{1cm}\code{\hspace{1cm}git checkout siscBlottnerSphere}
\end{enumerate}

\subsection{Code versions used for \S~4.2}
The version of \pressio~and \pressioFpy~used to generate
the results in \S~4.2 can be obtained in a similar way as follows:
\begin{enumerate}
\item Clone the core C++ \pressio~repository:\\
  \hspace{1cm}\code{\hspace{1cm}git clone https://github.com/Pressio/pressio.git}\\
  \hspace{1cm}\code{\hspace{1cm}git checkout siscBurgers1d}
\smallskip
\item Clone \pressioFpy:\\
  \hspace{1cm}\code{\hspace{1cm}git clone https://github.com/Pressio/pressio4py.git}\\
  \hspace{1cm}\code{\hspace{1cm}git checkout siscBurgers1d}
\smallskip
\item Clone the code for Burgers1d:\\
  \hspace{1cm}\code{\hspace{1cm}git clone https://github.com/Pressio/pressio-sisc-burgers1d.git}
  \smallskip

\end{enumerate}